\documentclass[letterpaper,twocolumn,10pt]{article}
\usepackage{usenix,epsfig,endnotes}
\usepackage{caption}
\usepackage{subcaption}
\usepackage{multirow}
\usepackage{amsmath}

\usepackage{xcolor}
\newif\ifcomment
\commenttrue
\commentfalse
\ifcomment
\newcommand{\shirin}[1]{{\bf \textcolor{purple}{Shirin: #1}}}
\newcommand{\sayak}[1]{{\bf \textcolor{red}{Sayak: #1}}}

\else
\newcommand{\shirin}[1]{}
\newcommand{\sayak}[1]{}
\fi

\begin{document}

\date{}

\title{\Large \bf A Large-Scale Analysis of Phishing Websites Hosted on Free Web Hosting Domains
}


\author{
{\rm Sayak Saha Roy}\\
University of Texas at Arlington \\
sayak.saharoy@mavs.uta.edu
\and
{\rm Unique Karanjit}\\
University of Texas at Arlington \\
unique.karanjit@mavs.uta.edu
\and
{\rm Shirin Nilizadeh}\\
University of Texas at Arlington \\
shirin.nilizadeh@uta.edu
}

\maketitle

\thispagestyle{empty}

\subsection*{Abstract}

Free Website Building services (FWBs) provide individuals with a cost-effective and convenient way to create a website without requiring advanced technical knowledge or coding skills. However, malicious actors often abuse these services to host phishing websites. In this work, we propose FreePhish, a scalable framework to continuously identify phishing websites that are created using FWBs. Using FreePhish, we were able to detect and characterize more than 31.4K phishing URLs that were created using 17 unique free website builder services and shared on Twitter and Facebook over a period of six months. We find that FWBs provide attackers with several features that make it easier to create and maintain phishing websites at scale while simultaneously evading anti-phishing countermeasures. Our study indicates that anti-phishing blocklists and browser protection tools have significantly lower coverage and high detection time against FWB phishing attacks when compared to regular (self-hosted) phishing websites. While our prompt disclosure of these attacks helped some FWBs to remove these attacks, we found several others who were slow at removal or did not remove them outright, with the same also being true for Twitter and Facebook. Finally, we also provide FreePhish as a free Chromium web extension that can be utilized to prevent end-users from accessing potential FHD-based phishing attacks.

\section{Introduction}

Phishing attacks are responsible for countless data breach incidents and credential fraud, leading to millions of dollars in financial damage~\cite{ciscotrend,cnbc100million:2019,secmag:2021}. 
The success of phishing scams is largely dependent on effective social engineering~\cite{downs2007behavioral}, which necessitates the creation of websites that convincingly mimic their legitimate counterparts. This task demands a good understanding of web design, attention to detail, and proficiency in mimicking trusted entities. While phishing kits~\cite{phishingkit2022} have emerged to streamline this process, the more effective ones often come with a hefty price tag. Furthermore, the security landscape has evolved, with recent research focused on detecting these phishing kits and the websites they generate~\cite{oest2018inside,bijmans2021catching,han2016phisheye,cui2021proactive}, rendering them less of a fail-safe approach for generating phishing scams.
In addition, the proficiency of anti-phishing entities in detecting regular phishing attacks has notably increased~\cite{oest2020phishtime,oest2020sunrise}. Therefore, attackers continuously innovate and unearth newer exploits~\cite{oest2019phishfarm,zhang2021crawlphish,song2021advanced} to evade detection. 
For example, they often invest significantly in procuring domains to host their scams, which is especially burdensome considering the frequent takedowns~\cite{oest2020sunrise} enacted by anti-phishing entities, leading to a relentless cycle of buying new domains to keep their operations active~\cite{bulkdomains}.
Moreover, to enhance the perceived legitimacy of the phishing sites, premium top-level domains (TLDs), such as .com~\cite{bleepingcomputer_tld} are often used.

Our research closely studies a trend for generating phishing attacks using 17 unique \textbf{Free Website Building services (FWBs)}. These services provide a platform that allows hosting websites at no cost and utilize user-friendly drag-and-drop builder interfaces which facilitates creating phishing attacks. These phishing websites 
are indistinguishable from those generated through manual coding or phishing kits.
Also, these FWB services provide several features that allow the creation of phishing scams that can evade anti-phishing detection mechanisms and resist takedowns for extended periods. 
In summary, using FWBs alleviates many of the challenges in maintaining and scaling phishing operations.

In this work, we propose \textit{FreePhish}, a framework for automatically identifying phishing attacks created using FWBs and shared on social media services, monitoring the reaction of anti-phishing entities against them, and reporting them for removal. Using FreePhish, we characterize the features and effectiveness of over 31.4K such phishing URLs that are shared on Twitter and Facebook over a period of six months.

Despite the widespread use of FWBs to host phishing scams, so far, there has been a limited understanding of these attacks due to only sporadic (and short) reports~\cite{bleepingcomp,infoworld,thesecurityledger} dedicated to this issue. In response to this knowledge gap, our work pioneers a large-scale, longitudinal study to meticulously characterize and detect these attacks.

The paper is structured as follows: We first establish the need to study these attacks by identifying their pervasiveness on both Twitter and Facebook over the period of two years in Section~\ref{pervasiveness}, followed by qualitatively studying these websites in Section~\ref{characterization} to find unique structural characteristics which distinguish them from regular phishing websites (i.e., those which are self-hosted), and make them more evasive to anti-phishing detection. We utilize these findings to develop FreePhish, in Section~\ref{freephish-framework}, our automated framework that continuously identifies FWB phishing threats in real-time, reports them to the respective FWBs and social media platforms for removal, and measures the effectiveness of anti-scam entities against them. FreePhish helped us identify more than 31.4K FWB phishing URLs from November 2022 to March 2023, and we dedicate Section~\ref{measurement} to closely investigate how well they are detected by
four popular phishing blocklists: PhishTank~\cite{phishtank}, OpenPhish~\cite{openphish}, Google Safe Browsing~\cite{safebrowsing}, and APWG eCrimeX~\cite{ecrimex:2022}, 76 security tools, as well as the FWB service providers and the social media networks (Twitter and Facebook). 
Our comprehensive measurement highlights several gaps in the prevalent anti-phishing ecosystem against FWB-based phishing attacks.
Finally, in Section~\ref{evasive-fhd}, we study how attackers further improve their efforts to host more evasive attacks on certain FWB domains, such as using i-frames, multi-step phishing attacks, and malicious drive-by downloads. Moreover, we introduce FreePhish as a Chromium web extension that can proactively prevent users from accessing these attacks.  
The primary contributions of our work are: 
\begin{enumerate} 
\item We have performed the first extensive analysis of phishing websites created using 17 free website creation services (FWBs), revealing features that enable evasion of common anti-phishing methods and website takedown resistance.

\item We created FreePhish, a machine learning (ML) framework capable of detecting FWB phishing websites. It auto-identifies, reports, and evaluates these websites, outperforming other ML anti-phishing measures with a 97\% accuracy.

\item Our framework detected 31.4K new FWB phishing attacks within six months. We assessed the effectiveness of popular phishing blocklists, browser protection tools, social media platforms and the FWBs themselves against these attacks. Our findings show low coverage and slow response time across all these attacks, highlighting significant gaps against these attacks in the present anti-phishing ecosystem.
\item
\urlstyle{tt}
We provide FreePhish as a free web extension that can be used in any Chromium based web-browser to proactively prevent users from accessing FWB phishing attacks, as well as open-source our dataset at \url{https://github.com/UTA-SPRLab/freephish/}
\urlstyle{rm}


\end{enumerate} 

 \section{Historical Pervasiveness of FWB Phishing Attacks}
\label{pervasiveness}
\begin{figure}[!htb]
\includegraphics[width=1\columnwidth]{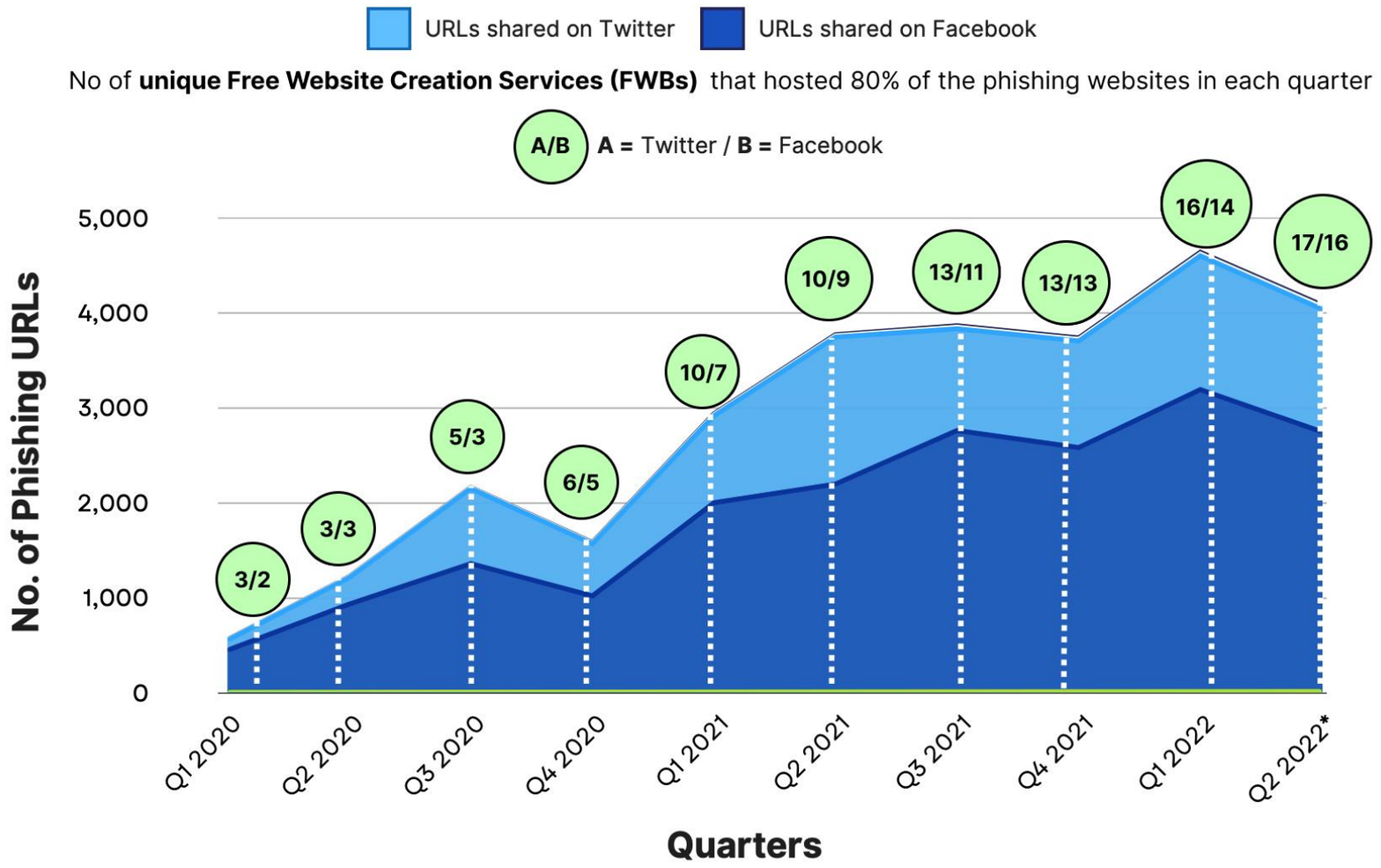}
\caption{Distribution of FWB phishing attacks shared on Twitter and Facebook from Jan. 2020 to Aug. 2022.}
\label{fhd-over-time}
\end{figure}
\begin{figure*}[!htb]
\begin{center}
\includegraphics[width=1\textwidth]{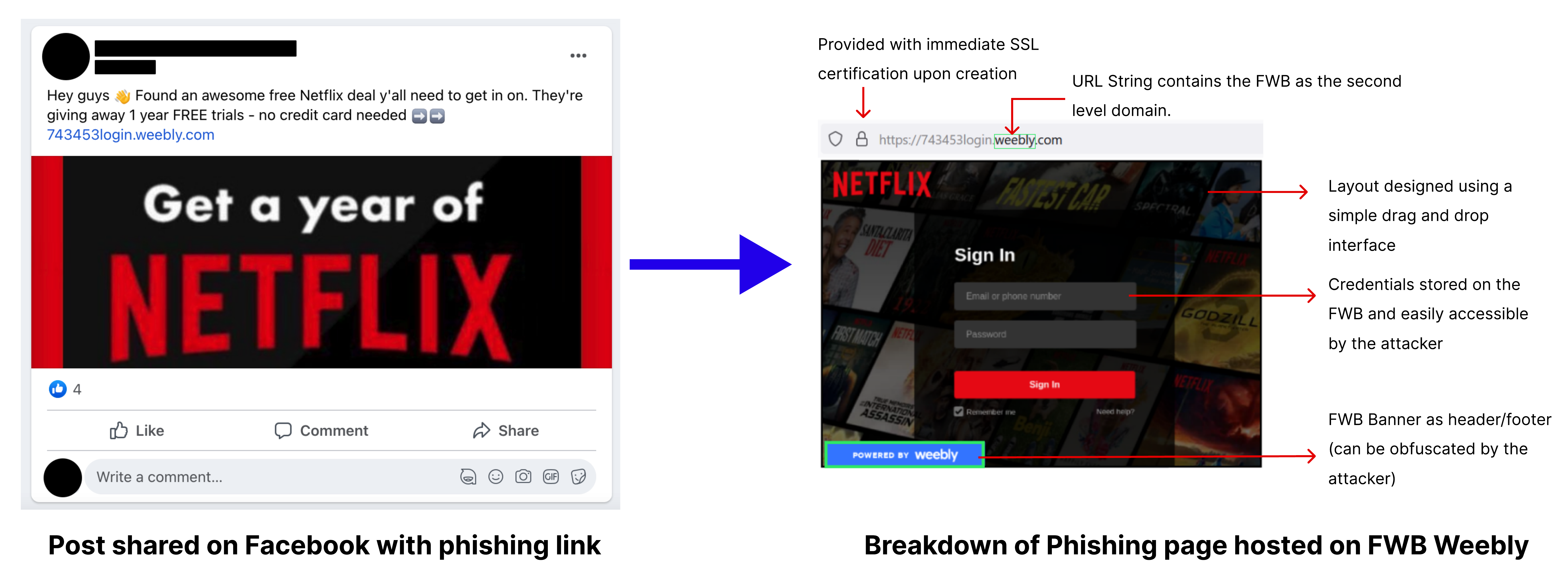}
 \caption{Example of a phishing website created on Weebly, a FWB service, and shared on Facebook. Note - This website has already been taken down, and thus the full URL string is not obfuscated from the image.}
\label{fwd-examples}
\end{center}
\end{figure*}
To underscore the necessity of studying phishing websites created using free website builder services, we documented a rising trend in these attacks over a two-year period. Using the official Twitter API~\cite{Twitter:2020} and Meta's CrowdTangle~\cite{crowdtangle}, we compiled 3.1 million URLs from Twitter and 1.4 million URLs from Facebook that contained distinct second-level domains in their URL strings over the period of January 2020 to August 2022. This decision to only collect URLs with a second-level domain was driven by the intention to analyze websites that are created under another domain. 
For instance, in the URL \texttt{mywebsite.000webhost.com}, 000webhost serves as the second-level domain.
The URLs were then scanned using VirusTotal~\cite{VirusTotalAPI:2020}, a service that collates detection scores from 80 different anti-phishing tools. URLs with two or more detections were labeled as \texttt{phishing}, conforming to standards set in previous literature for such identification~\cite{oprea2018made,sharif2018predicting,vtpaper_blackbox}. In addition to recognizing potential FWB phishing attacks over time, this analysis also helped us identify the FWBs that are being abused by attackers.
We labeled 34.7K unique URLs as phishing, out of which 25.2K URLs (16.3k URLs originating from Twitter and 8.9K URLs from Facebook) utilized 17 unique free website-building services. The remaining URLs belonged to Dynamic DNS services (such as DuckDNS, Netlify etc.) which is outside the purview of our study.
We refer to these 25.2k URLs as the \textit{initial dataset} or \texttt{D1} throughout the rest of the paper. 
Figure~\ref{fhd-over-time} displays the distribution of these attacks over time across both social media platforms.  Not only does this data highlight a marked escalation in the quarterly frequency of such attacks, but it also reveals a strategic shift in attackers' preference toward adopting newer hosting services. For each analyzed month, we identify the unique domains that accounted for 80\% of the FWB phishing attacks during that period. It is important to acknowledge that the actual scale of FWB phishing attacks is likely larger than our dataset suggests, given the proactive measures both platforms employ to flag or remove malicious content~\cite{TwitterSafety2023,FacebookTransparency2023}.
Our conclusions are further substantiated by Interisle's in-depth studies conducted between 2020 and 2023~\cite{interisle_2023,interisle_2022,interisle_2021,interisle_2020}. We draw particular attention to their analyses concerning the abuse of Subdomain Providers, which include not just FWB services but also Dynamic DNS vendors like DuckDNS and Netlify$—$although the latter are outside the purview of our research. Interisle's longitudinal reports corroborate two primary trends: firstly, there is an ascending trajectory in the utilization of FWB services for orchestrating phishing attacks; secondly, attackers are progressively exploiting newer FWB platforms, thereby mirroring the trends observed in our own study.

It is crucial to highlight that, due to the age of these URLs (some of these URLs are over two years old), we have refrained from utilizing them for any further longitudinal studies, for example, understanding the effectiveness of anti-phishing measures against FWB attacks. Instead, we utilize the zero-day FWB phishing attacks identified by our FreePhish framework in Section~\ref{measurement}. 
We also reported these 25.2k URLs to Twitter, Facebook, and the FWBs directly, thereby facilitating their removal from online platforms.

%

\section{Characterizing FWB Phishing Attacks}
\label{characterization}
\begin{figure*}[!htb]
\begin{center} 
\includegraphics[width=0.85\textwidth]{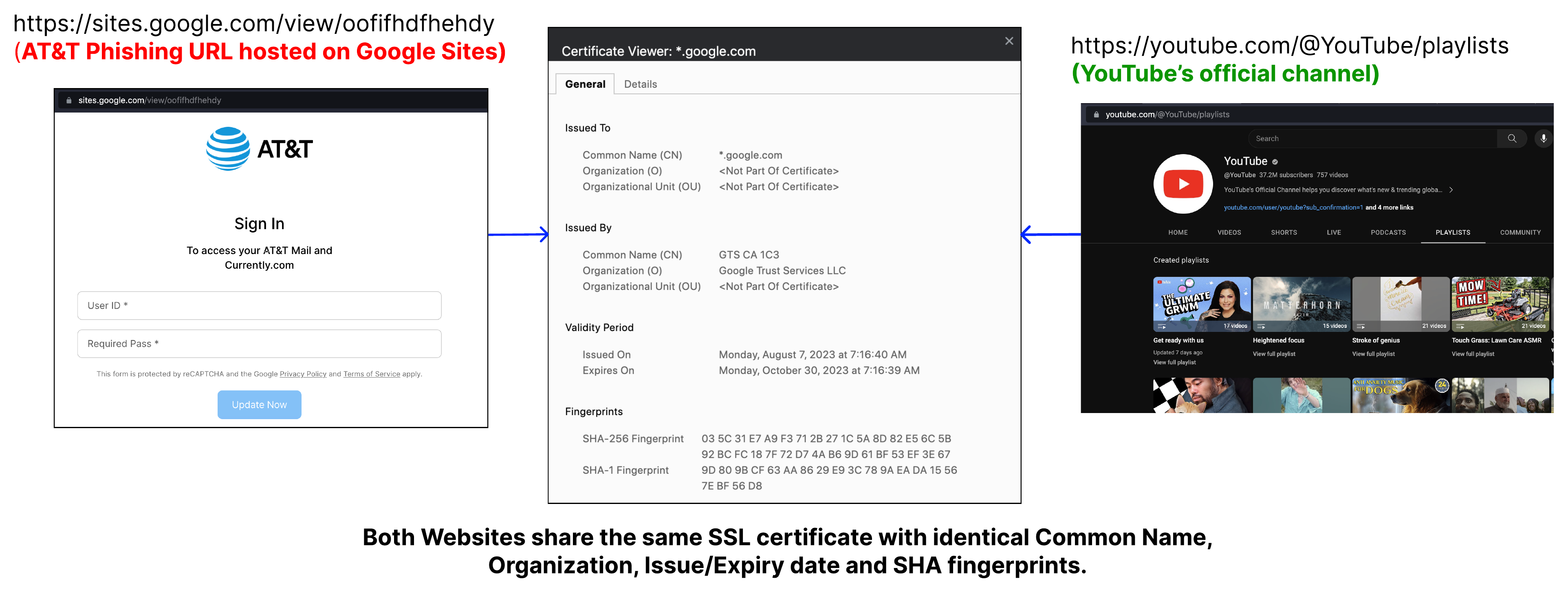}
\caption{Example shows FWB websites created using Google Sites share the same certification as YouTube. Note - The phishing website has already been taken down, and thus the full URL string is not obfuscated from the image.}
  \label{fig:share-ssl}
\end{center}
\end{figure*}

Having established the prevalence of phishing attacks originating from Free Website Builders (FWBs), our next objective was to understand the unique characteristics of these attacks. To do so, we carefully examined a random sample of 5K URLs from Dataset D1 to highlight the specific features that make FWB-created phishing attacks particularly advantageous to the attacker. Figure~\ref{fwd-examples} illustrates the anatomy of a phishing attack created on Weebly, a popular FWB and imitating the HTML login page.
Two university students, both specializing in Computer Security with a solid understanding of social engineering scams, were chosen as coders for this qualitative evaluation. 
The coders primarily assessed whether the URL and website appearance attempted to spoof one of 409 brands targeted by phishing attacks, as reported by OpenPhish in August 2022{~\cite{openphish_brands}}. In addition, the coders evaluated whether the websites contained text fields designed for gathering sensitive information (such as email addresses, passwords, SSN etc). For websites that did not meet these criteria but initiated a file download, the coders conducted additional scans using VirusTotal and marked the files as malicious if they generated four or more detections. Prior literature has established this detection threshold as being sufficient for identifying malware samples~\cite{oprea2018made,sharif2018predicting}.

Of the 5K randomly sampled URLs, 4,656 were confirmed to be phishing. The initial Cohen's Kappa inter-rater agreement was 0.78 (i.e., high agreement). 
Disagreements primarily occurred due to: misinterpretation of brand spoofing (i.e., differing views on how effectively a website mimics a legitimate brand), evasive attacks (Coder \#1 failed to recognize two-step phishing attacks as harmful), assessment of text fields (Coder \#1 overlooked the relevance of address fields and phone numbers in indicating phishing intent) and language representation (Coder \#2 was unable to identify phishing intentions in eight websites that primarily used Spanish and Chinese languages).
All disagreements were resolved through discussion and consensus. 
We further analyzed a sample of the confirmed true positive set and evaluated the design interfaces of the 17 unique FWB services to identify several factors contributing to the convenience and efficiency of launching phishing attacks using FWB services. 
We will discuss these findings in detail in the following paragraphs.

\textbf{Initial Investment:}
\label{investment}
Conventionally, phishing attacks have a low life span, getting detected by anti-phishing providers and subsequently removed by the hosting domain within a few hours~\cite{oest2020sunrise}. 
In response to the countermeasures, attackers frequently redeploy phishing attacks across multiple domains, which need ongoing investments~\cite{oest2020sunrise,oest2020phishtime}.
However, free website builders have turned this situation in the attacker's favor, as they allow users to create and host websites on their servers for free, thus allowing a straightforward way to scale attacks. Moreover, if they get blocklisted on one FWB, they can switch to another without any added expenditure. 

\textbf{Easy of creation:} 
\label{ease-of-creation}
Phishing attacks are designed to imitate legitimate websites closely, deceiving users into sharing their sensitive information. Designing the webpage and its related components requires a level of technical expertise.
While dedicated phishing kits exist to aid attackers in designing webpages, a recent focus on detecting such kits has lowered the potency of such attacks.
On the other hand, FWBs provide easy-to-use drag-and-drop website creation interfaces. These builders do not require any coding experience and can be used to create and publish phishing websites in a matter of minutes. They also handle data entered in the credential fields, provide additional applications like SEO boosting and social media integration, and allow the embedding of custom HTML code, which can be exploited by attackers. 

\textbf{Immediate SSL Certification:}
\label{ssl-cert}
Users are often advised to check if a website has SSL certification to avoid phishing scams~\cite{firefoxssl}. However, SSL based phishing attacks are widespread~\cite{drury2019certified}, with more than 49\% of phishing URLs now having SSL certification~\cite{halfsslphishing}. The registration of a new phishing website typically involves a certification payment and a verification process that can up to several days~\cite{sslverification}. 
More affordable or even free options like Let's Encrypt~\cite{lets_encrypt} and ZeroSSL~\cite{zerossl} offer Domain Validation (DV) certificates with a 90-day validity period. While these DV certificates are easier to obtain, they demand additional implementation effort and are generally considered less trustworthy than Extended Validation (EV) or Organization Validation (OV) certificates~\cite{ssl_ev,comodo_ev}. The latter types require website owners to undergo a more rigorous verification process, which in turn enhances their credibility.
A study by PhishLabs~\cite{phishlabs_ev} reveals that most SSL-based phishing sites use DV certificates. They also point out an increasing trend of attackers adopting EV certificates to make their fraudulent sites appear more legitimate.
In contrast, FWB Services automatically provides all new websites with either an EV or OV certificate, eliminating the need for additional verification steps and easing the implementation process.
Figure~\ref{fig:share-ssl} shows one example where a phishing attack created on FWB service Google Sites has the same SSL certificate as YouTube.com, an official Google platform. 
Moreover, considering that ML-based phishing detection models often positively co-relate SSL certification to the website being benign~\cite{afroz2011phishzoo,roopak2019effectiveness}, having this feature in FWB phishing attacks makes it easier to evade detection.

\textbf{Premium TLDs:}  
\label{premium-tlds}
Domain providers price their domains based on the popularity of the top-level domains (TLDs). 
For example, the \textit{.com, .org} TLDs are more expensive than the TLDs like  \textit{.xyz, .live}. 
To save on investment, attackers usually launch phishing websites on cheaper TLDs~\cite{abusedtlds:2022}, which have, in turn, tuned blocklists, tools, and automated machine learning tools to use TLDs as a heuristic for detection~\cite{xiang2011cantina+,thakur2014catching,prakash2010phishnet}.
Recent studies have found that users have higher trust in \emph{.com} TLDs ~\cite{varn:2022,growthbadger_domain}.
All 14 out of 17 FWBs provide users with a \textit{.com} TLD for their websites, thus allowing the attackers to get a premium TLD at no cost. 
Out of the 4,656 domains that our coders identified as phishing sites, a staggering 4,139 $—$ or approximately 89\% were hosted on the 14 FWB services that offer a .com Top-Level Domain (TLD). 
Thus, along with not appearing in certification logs, this feature can further aid in evading user suspicion and anti-phishing detection.

\textbf{Longer Domain Age:}  
\label{longer-domain-age}
Phishing websites usually remain active for a very short period of time since they are either blocklisted/taken down or abandoned by the attacker.
Thus several anti-phishing tools consider the \textit{domain age} to be an important heuristic feature~\cite{patil2019methodical,oest2018inside,sanglerdsinlapachai2010using}. 
However, phishing attacks created using FWBs are essentially sub-domains of the FWB itself, and thus have the same domain age as the FWB, as recorded by registrar information databases such as WHOIS~\cite{whois}.
Since the FWBs (e.g., Weebly, Google Sites, Squarespace) are several years old, detection models which utilize domain age as an important heuristic, might associate FWB-hosted phishing attacks to be legitimate. 
The median domain age of the 4,656 URLs in our dataset labeled as phishing was 13.7 years, according to their WHOIS domain record.
On the other hand, the same number of self-hosted phishing URLs collected from PhishTank~\cite{phishtank}, an open-source anti-phishing blocklist, had a median age of only 71 days.

\textbf{High Code Similarity Between Legitimate and Phishing Websites:} 
\label{website-code-similarity}
Automated phishing detection measures often assume that the code structure of phishing websites differs significantly from legitimate websites~\cite{alkhozae2011phishing,opara2020htmlphish}. Moreover, some detection methods tend to compare the source code of new websites with that of known phishing websites to identify potential threats~\cite{roopak2014novel}.
However, this approach could become less effective in cases where free website builders (FWBs) are used. These platforms allow users to create websites using predefined templates, leading to considerable similarities between the HTML code of different websites hosted on the same service, including both legitimate and phishing sites.
In our study, we used the Levenshtein edit distance algorithm, a popular metric for computing similarity between text sequences~\cite{alvarez2007using,rani2017enhancing}, to measure the extent of similarity in HTML code blocks between benign and phishing websites hosted on these FWBs. The detailed methodology we used for this computation is explained in Appendix~\ref{appendix-code-similarity}.
Table~\ref{table-website-similarity} illustrates the median code similarity between phishing and benign websites created using some popular FWBs, showing a high median code similarity for several FWBs. For example, websites created on Weebly have a median similarity of 79.4\%. This indicates that attackers can potentially use predefined templates provided by FWBs to make their phishing websites appear more like legitimate websites, thus increasing the chances of evading detection. 
\begin{table}[]
\centering
 \resizebox{0.7\columnwidth}{!}{%
\begin{tabular}{|l|c|c|}
\hline
FWB & \# URLs & Median similarity \\ \hline
Weebly & 6,309 & 79.4\% \\ 
000webhostapp & 3,714 & 68.1\% \\ 
Blogspot & 2,492 & 63.8\% \\ 
Google Sites & 1,547  & 72.4\% \\ 
Wix  & 1,801 & 63.7\% \\  
Github.io & 582 & 37.4\% \\
\hline
\end{tabular}}
\caption{Website code similarity between FWB phishing and benign websites.}
  \label{table-website-similarity}
\end{table}

\textbf{Increased Difficulty of Discovery:} 
\label{harder-to-discover}
Phishing websites created through FWBs are more challenging to discover for multiple reasons. Firstly, once domains obtain their SSL certification, they typically appear on the Certificate Transparency (CT) Log network~\cite{certificatetransparencynetwork:2022}. As many phishing domains now have SSL certification~\cite{halfsslphishing}, various anti-phishing methods utilize this log to identify new attacks~\cite{drichel2021finding,fasllija2019phish,sakurai2020discovering}. 
However, we found that all websites created using all 17 FWB services inherited the same SSL certification used by the service itself.  Notably, three of these services $-$ Weebly~\cite{weebly_cert}, Wix~\cite{wix_cert}, and Mailchimp~\cite{mailchimp_cert} highlight the fact that websites created for free through their platforms are automatically provided with complimentary SSL certification. Having the same SSL certification as their hosting domain renders FWB phishing attacks effectively invisible in Certificate Transparency (CT) logs as they do not appear as individually registered domains. 
Notably, the standard practice of certificate re-issuance for compromised or vulnerable domains~\cite{zhang2014analysis} does not apply to FWB-hosted websites, since they naturally inherit the SSL certificates from their primary FWB domain.
Secondly, numerous anti-phishing crawlers examine websites indexed by search engines to discover new phishing attacks~\cite{huh2011phishing,rao2019jail,gupta2020phishing}. However, only a small fraction (4.1\%) of the 25.2K FWB phishing URLs in our historical dataset were indexed by Google. Google does not index websites without any incoming links (a common occurrence for phishing attacks hosted under subdomains) or if the website source code contains a $<noindex>$ meta tag, instructing search engines not to index the website~\cite{googleindexsubdomain}. Among the FWB-hosted phishing URLs in our dataset, 44.7\% contained the $<noindex>$ meta-tag, further contributing to the difficulty of their discovery.

\section{FreePhish Framework} 
\label{freephish-framework}
\begin{figure*}[!htb]
\centerline{\includegraphics[width=1\textwidth]{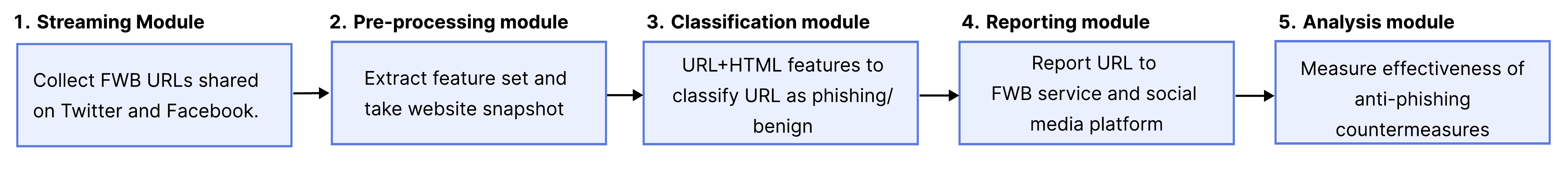}}
\caption{High-level overview of the FreePhish framework}
\end{figure*}


In this section, we detail the development of \textit{FreePhish}, a real-time framework designed to detect and combat FWB-based phishing attacks shared on Twitter and Facebook. FreePhish is not only capable of detecting and reporting these attacks to hosting providers in real-time but also longitudinally monitors the responses of anti-phishing entities against them.

Our framework, illustrated in Figure~\ref{freephish-framework}, comprises five core components: (1) \emph{Streaming module:} Operates at 10-minute intervals, monitoring for FWB URLs (whether phishing or non-phishing) shared on Twitter and Facebook. The module also captures the full snapshot of each identified website. (2) \emph{ Preprocessing module:} Saves a full snapshot of the website (including screenshots and source code) and extracts several URL and HTML-based features derived from our qualitative analysis in Section~\ref{characterization}. (3) \emph{Classification module:} Employs a Random Forest classifier to analyze the extracted features, predicting whether the website is engaging in a phishing attack.
(4) \emph{Reporting module:} Reports each URL flagged as a phishing site to the respective FWB service or hosting provider, and subsequently monitors their response.
(5) \emph{Analysis module:} Evaluates the effectiveness of anti-phishing entities, monitoring each URL at regular intervals.
We discuss each of these components in more detail.

\subsection{Streaming and Pre-processing Module}

The streaming module utilizes the Twitter and CrowdTangle APIs to collect new posts from Twitter and Facebook every 10 mins. It utilizes a regular expression to extract the URL from the post, which is followed by the pre-processing module. It stores a full snapshot of the website and extracts several URL-based, HTML-based, and FWB features, such as obfuscated banner code, credential fields, presence of \emph{<noindex>} tags, etc.  This feature set is passed to the Classification module.

\subsection{Classification Module}
Like most evasive phishing websites, FWB attacks share significant similarities with self-hosted phishing attacks but introduce subtle modifications to the website structure by using several features of FWBs, allowing them to bypass existing detection mechanisms. The motivation behind crafting this classification module is twofold. Firstly, we want to achieve a high success rate in detecting these attacks. Secondly, we want to ensure time efficiency. Since a large volume of websites is shared on social media throughout the day, a slower classification model can exponentially hamper the framework's overall performance.
To accomplish these objectives, we first build the ground-truth dataset. We test four state-of-the-art ML-based phishing detection models to identify the candidate which provides the optimal trade-off between performance and runtime efficiency. Finally, we augment the model by adding and removing FWB phishing-specific features (Section~\ref{feature-extraction}), then train and test the augmented model on our dataset and gauge its performance.   

\textbf{Ground truth collection:}  
\label{groundtruth}
In order to create our ground-truth dataset, we use the manually verified 4,656 true positive URLs in the initial dataset (D1). We also selected and manually verified an equal number of true negative benign FWB-created URLs from D1 that were shared on Twitter (n=3,299) and Facebook (n=1,357).

\textbf{Choosing the optimum model:} 
\begin{table*}[htb]
\centering
\resizebox{0.85\textwidth}{!}{%
\begin{tabular}{|c|c|c|c|c|c|c|}
\hline
\textbf{Model} & \textbf{Accuracy} & \textbf{Precision} & \textbf{Recall} & \textbf{F1-score} & \textbf{Total Time Taken (sec)} & \textbf{Median Runtime (sec)} \\
\hline
VisualPhishNet & 0.76 & 0.78 & 0.72 & 0.75 & 14,802 & 5.1 \\
PhishIntention & 0.96 & 0.98 & 0.94 & 0.96 & 32,958  & 11.3 \\
URLNet & 0.68 & 0.70 & 0.67 & 0.68 & 4,471 & 1.9 \\
Base StackModel & 0.88 & 0.89 & 0.87 & 0.88 & 8099.4 & 2.4 \\
\textbf{Our Model} & \textbf{0.97} & \textbf{0.96} & \textbf{0.97} & \textbf{0.96} & \textbf{8,658} & \textbf{2.8} \\
\hline
\end{tabular}}
\caption{Comparison of various phishing detection models.}
\label{tab:model_comparison1}
\end{table*}

We compared the performance of four state-of-the-art ML-based phishing detection models. 
These models included two that rely on the visual features of the website: VisualPhishNet~\cite{abdelnabi2020visualphishnet} and PhishIntention~\cite{liu2022inferring}, one that relies on both the URL string and  HTML representation of the website: StackModel~\cite{li2019stacking}, and one that relies on the semantic representation of the URL string only: URLNet~\cite{le2018urlnet}. All models were tested across our entire testing set. 
The models were tested on a system running on an Intel Xeon W Processor with 64GB of RAM and 2x NVIDIA RTX 2080Ti GPU.
Table~\ref{tab:model_comparison1} illustrates that URLNet has the lowest median runtime of 1.9 secs but also the lowest recall of 0.68. 
VisualPhishNet has a median runtime and a recall of 5.1 secs and 0.75, while StackModel has that of only 2.4 secs and 0.87, respectively.  
PhishIntention performed the best with a recall of 0.97, probably because, unlike the other models, it does not rely solely on visual features or URLs
, but incorporates both static and dynamic analyses of the website's entire workflow~\cite{liu2022inferring}. 
However, its complexity makes it far slower at classifying our samples, with a median runtime of 11.3 secs per URL, which can lead to significant overhead in our framework. 
We thus choose the StackModel since it strikes the best balance between efficiency and runtime, and we improve upon it for detecting FWB phishing attacks. 

\textbf{Feature extraction:}  
\label{feature-extraction} 
Our model builds on the feature set utilized by the StackModel, which includes 8 URL-based features (such as the usage of suspicious symbols, sensitive vocabulary in the URL string, similar brand names, the length of the URL, etc.) and 12 HTML-based features (such as the number of internal and external links, the presence of empty links, the inclusion of a login form, the length of HTML content). The full list of features can be found in the original publication by Li et al.~\cite{li2019stacking}. 
Two of the features StackModel uses to detect phishing websites, including the presence of 'https' and multiple top-level domains, do not apply to those created by FWBs, because all FWB websites utilize 'https' and FWB phishing attacks typically contain only one TLD. 
Therefore, we did not use them in our model. 

In their place, we introduced two new features specifically tailored for FWB phishing detection, drawing on our analysis as discussed in Section~\ref{characterization} - 
\emph{Obfuscating FWB Footer:} This feature recognizes the trend among FWB services to include a header or footer banner on their free websites. To make their fraudulent sites appear more legitimate, phishers often alter the source code of these websites to hide these banners. They might, for example, add a \textit{$<visibility: hidden>$} parameter to the $<div>$ tag containing the banner code.
\emph{Preventing Indexing:} As detailed in Section~\ref{harder-to-discover}, attackers frequently employ \textit{noindex} meta-tags to prevent search engines from indexing their URLs, which indirectly reduces traffic from anti-phishing crawlers. We included a feature that checks for the presence of these tags in the website's source code.

\textbf{Model training and performance:}  
We employ the methodology utilized by Li et al.~\cite{li2019stacking} for training the StackModel architecture. To summarize, their approach utilizes a two-layer stacking model that combines multiple ML models. The models employed in each layer include Gradient Boosting Decision Tree (GBDT), XGBoost, and LightGBM.
A strategy similar to K-fold cross-validation is employed for training, where for each iteration, the dataset (n=4,656 phishing samples, 4,656 benign samples) is partitioned into 70\% for the training set and 30\% for the testing set. This process continues until each basic model has predicted all samples. The initial layer utilizes both original features and prediction results derived from majority voting as inputs for the second layer. The second layer combines the input features with its output results, and these composite features are then used to train a final GBDT model. This model then predicts which webpages are phishing sites. The proposed model utilizes both URL and HTML-based features for prediction highlighted in Section~\ref{feature-extraction}. As illustrated in the last row of Table~\ref{tab:model_comparison1}, our augmented StackModel has an F1-score of 0.96 and a median run-time of only 2.8 secs.
\subsection{Reporting Module}
URLs identified by the classification model (as phishing) are reported immediately to the respective FWB service and social media platforms (i.e. Twitter and Facebook). This ensures the timely and ethical disclosure of identified threats from our platform to the sources sharing them online. We used Python Selenium to automate the submission of reports through the Chromium Web Browser, as neither of the platforms have the feature to report malicious URLs through API.
According to prior literature, including evidence-based information like website screenshots expedites the malicious URLs evaluation process~\cite{oest2020phishtime}. As a result, our FreePhish reported URLs to include the full URL, a screenshot of the site, and the targeted organization's name. For safety reasons, our experiment underwent a thorough review and was approved by our organization's Office of Information Security before we submitted the phishing reports. We explore our findings regarding the effectiveness of FWB and social media platforms in removing these threats in Section~\ref{measurement}.
Contrarily, we refrain from reporting to blocklists. Community-based blocklists, such as APWG eCrimeX and PhishTank, list reported URLs in their feed immediately without any verification. This could undermine our longitudinal measurement's validity. This also applies to several tools we studied for browser protection, which rely on community threat intelligence.

\subsection{Analysis Module}
We conduct a longitudinal evaluation to assess the effectiveness of various anti-phishing entities against the discovered FWB phishing attacks.
Two key performance indicators were considered: \textit{coverage} - the proportion of FWB-hosted phishing URLs detected or removed within one week of appearing in our dataset, and \textit{response time} - the duration from a URL's first appearance in our dataset to its detection or removal by the entity.
The proceeding paragraphs describe the analyses: 

\textbf{Anti-phishing blocklists:}
For each FWB phishing URL, we checked if it was present on four anti-phishing blocklists: Google Safe Browsing (GSB), PhishTank, OpenPhish, and APWG eCrimeX, using their respective APIs, at regular intervals of 10 minutes till the URL became inactive. 
GSB is regularly used on several browsers running on the Chromium engine, such as Google Chrome ~\cite{chromesafebrowsing}, Firefox ~\cite{firefox:2021}, Vivaldi~\cite{vivaldisafebrowsing}, etc. PhishTank and OpenPhish regularly contribute their data to several anti-phishing tools and web browsers ~\cite{phishfriends}. 
APWG's eCrimeX also shares its data with several organizational anti-phishing entities~\cite{oest2020phishtime}.

\textbf{Browser protection tool analysis:} We used VirusTotal~\cite{virustotal} to scan both FWB-based and self-hosted phishing URLs every 10 minutes, from when the URL first appeared in our dataset, up till when it became inactive, or up to a week, whichever was sooner. 
This is to mitigate the possibility of outdated scores~\cite{vtpaper_blackbox}. 
Unlike our previous analysis, where we evaluated the performance of each blocklist separately, we aggregated the VirusTotal scores, which provided the detection rate for up to 76 different anti-phishing engines and blocklists. VirusTotal data also includes GSB, PhishTank, and OpenPhish, and we excluded the scores of these blocklists to avoid redundancy with our previous analysis. 

\textbf{FWB response analysis:}   
The most efficient way to prevent users from falling victim to phishing attacks is if the website gets removed by the hosting providers. 
To evaluate domain takedown of FWB-based phishing URLs, we checked if the website was active every 10 mins from when it first appeared in our dataset. 

\textbf{Social media platform response analysis:}  
Figure~\ref{fhd-over-time} illustrates how FWB-based  phishing attacks are becoming a persistently growing threat on Twitter and Facebook, with attackers exploiting more FWBs to host their attacks. 
After collecting the initial tweet (containing the phishing link), we used the Twitter Academic API~\cite{twitter_developer} to check if the tweet had been deleted at regular intervals of 10 minutes. 
To check the same for Facebook posts, we check the URL of each post, identifiable by their unique post id.  

\section{Measurement Results}
\label{measurement}
\begin{figure}[t]
\centering
  \includegraphics[width=0.8\columnwidth]{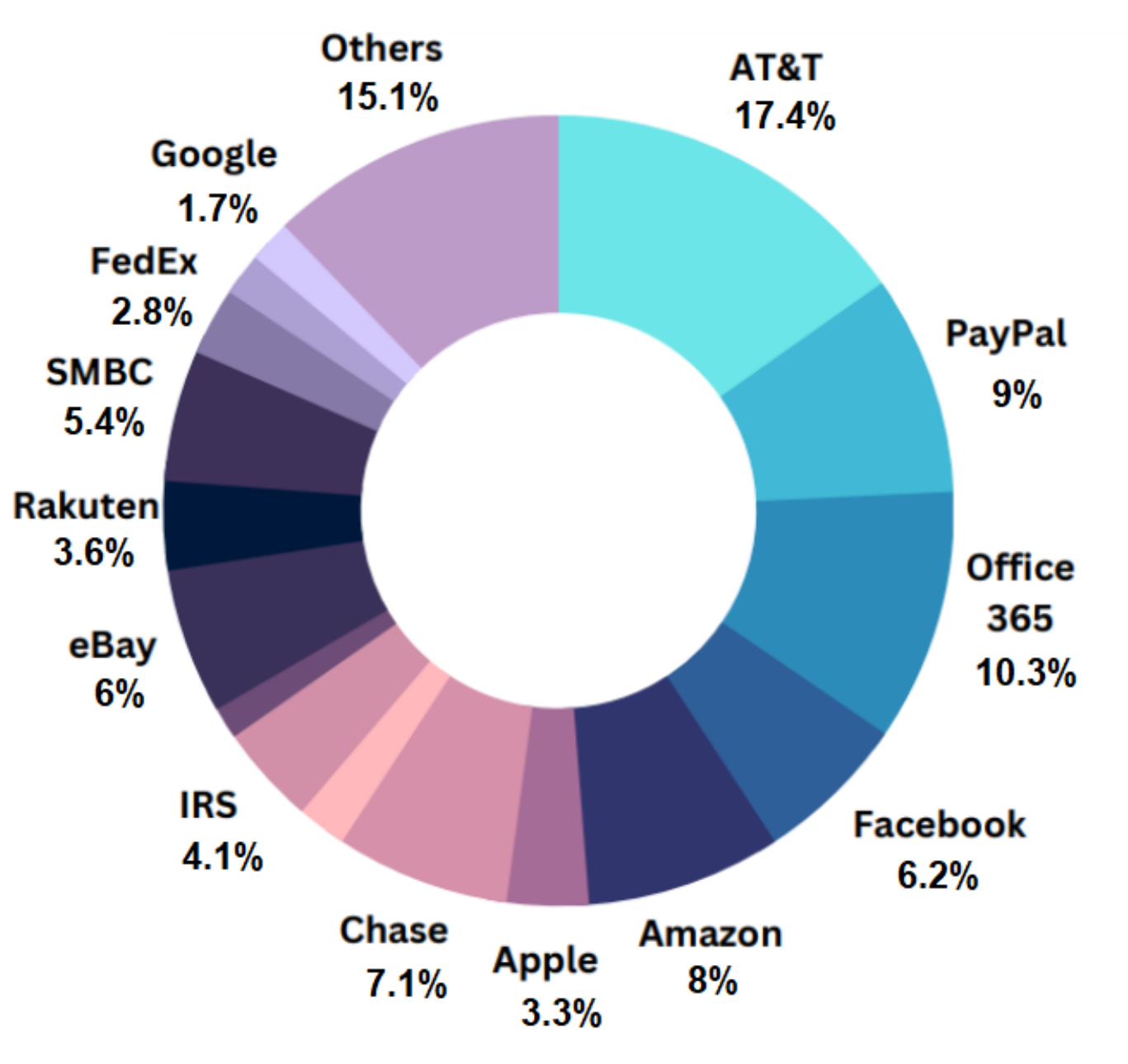}
\caption{Targetted Organizations}
\label{targets}
\end{figure}

We ran FreePhish for a period of six months, from November 2022 to May 2023, identifying 31,405 zero-day phishing attacks (19,724 URLs from Twitter, 11,681 URLs from Facebook), and  created using free website builders. These attacks targeted 109 unique brands and organizations, with the most frequently imitated ones illustrated in Figure~\ref{targets}. In this section, we identify the effectiveness of different phishing countermeasures against these attacks, including Blocklists, Browser protection tools, the respective FWB services themselves, as well as the social media platforms - Twitter and Facebook. 

\textbf{Comparison with self hosted phishing attacks:} For each analysis, we also compared the detection coverage and response times for FWB-hosted phishing attacks with that of phishing attacks hosted on self-hosted domains. To find these self-hosted phishing attacks, we ran the base StackModel over a daily stream of Twitter and Facebook posts collected using their respective APIs from November 2022 to May 2023 (The same time period as our measurement for the FWB phishing attacks). We selected a random sample of 31,405 URLs and ensured that the distribution of the sample matched the distribution of the FWB phishing attacks across the two platforms: 19,724 phishing URLs from Twitter and 11,681 phishing URLs from Facebook.

\label{blocklist-results}

\subsection{Blocklist Performance}

\begin{table*}[]
\centering
\caption{Blocklisting performance and response time of anti-phishing entities against FWB phishing attacks.
}
\resizebox{0.85\textwidth}{!}{%
\begin{tabular}{|c|ccc|ccc|}
\hline
\multirow{2}{*}{\textbf{Method}} & \multicolumn{3}{c|}{\textbf{FWB Phishing websites}} & \multicolumn{3}{c|}{\textbf{Self-hosted Phishing Attacks}} \\ \cline{2-7} 
 & \multicolumn{1}{c|}{\textbf{Coverage}} & \multicolumn{1}{c|}{\textbf{Min/Max (hh:mm)}} & \textbf{Median Speed (hh:mm)} & \multicolumn{1}{c|}{\textbf{Coverage}} & \multicolumn{1}{c|}{\textbf{Min/Max}} & \textbf{Median Speed} \\ \hline
PhishTank & \multicolumn{1}{c|}{4.08\%} & \multicolumn{1}{c|}{0.10/116:13} & 07:11 & \multicolumn{1}{c|}{17.4\%} & \multicolumn{1}{c|}{0:03/122:03} & 02:30 \\ 
OpenPhish & \multicolumn{1}{c|}{11.70\%} & \multicolumn{1}{c|}{0:02/191:30} & 13:20 & \multicolumn{1}{c|}{30.5\%} & \multicolumn{1}{c|}{0:01/146:18} & 02:21 \\ 
GSB & \multicolumn{1}{c|}{18.44\%} & \multicolumn{1}{c|}{0:02/148:05} & 06:01 & \multicolumn{1}{c|}{74.2\%} & \multicolumn{1}{c|}{0:01/146:26} & 00:51 \\ 
eCrimeX & \multicolumn{1}{c|}{32.90\%} & \multicolumn{1}{c|}{0:07/137:43} & 08:54 & \multicolumn{1}{c|}{47.9\%} & \multicolumn{1}{c|}{0:04/133:05} & 04:26 \\ 
Social media Platform & \multicolumn{1}{c|}{23.06\%} & \multicolumn{1}{c|}{0:04/125:53} & 10:25 & \multicolumn{1}{c|}{50.9\%} & \multicolumn{1}{c|}{0:07/114:23} & 03:41 \\ 
Hosting domain & \multicolumn{1}{c|}{29.38\%} & \multicolumn{1}{c|}{0:19/158:25} & 09:43 & \multicolumn{1}{c|}{77.50\%} & \multicolumn{1}{c|}{0:08/135:29} & 03:47 \\ \hline
\end{tabular}}
\label{table-fhd-overall}
\end{table*}

\begin{figure}[t]
\centering
  \includegraphics[width=1\columnwidth]{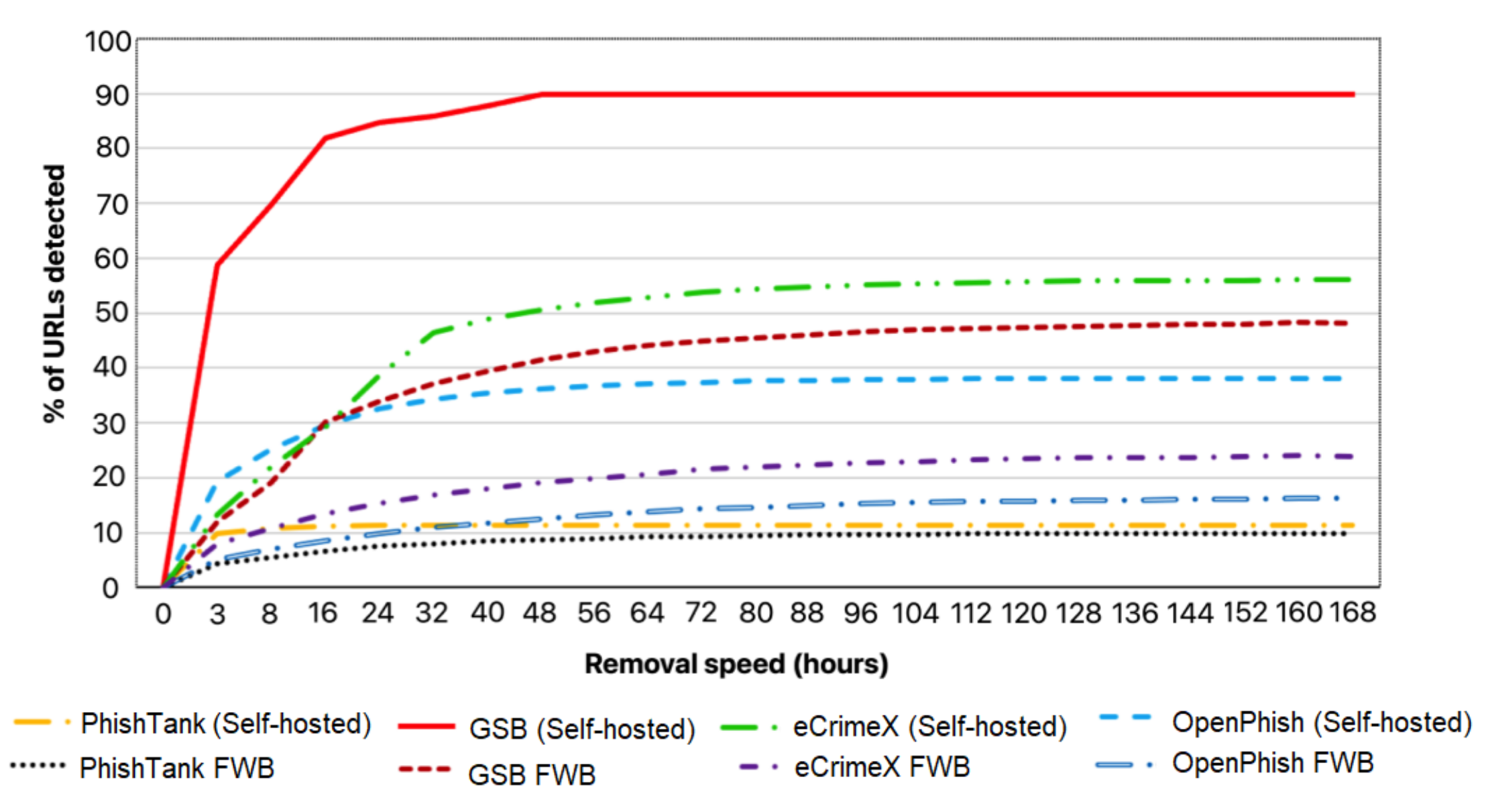}
\caption{Coverage and speed of blocklists against FWB and self-hosted phishing attacks}
  \label{fig:blocklist_blocking_overtime}
\end{figure}

\begin{table*}[]
\caption{Coverage and response times of different countermeasures against FWB phishing attacks.}
\label{table-fhd-per-domain}
\resizebox{\textwidth}{!}{%
\begin{tabular}{|c|c|cc|cc|cc|cc|cc|cc|}
\hline
\multirow{2}{*}{\textbf{Domains}} & \multirow{2}{*}{\textbf{URLs}} & \multicolumn{2}{c|}{\textbf{Domain}} & \multicolumn{2}{c|}{\textbf{Platform}} & \multicolumn{2}{c|}{\textbf{PhishTank}} & \multicolumn{2}{c|}{\textbf{OpenPhish}} & \multicolumn{2}{c|}{\textbf{GSB}} & \multicolumn{2}{c|}{\textbf{eCrimeX}} \\ \cline{3-14} 
 &  & \multicolumn{1}{c|}{\textbf{Removal Rate}} & \textbf{Speed (hh:mm)} & \multicolumn{1}{c|}{\textbf{Removal Rate}} & \textbf{Speed} & \multicolumn{1}{c|}{\textbf{Removal Rate}} & \textbf{Speed} & \multicolumn{1}{c|}{\textbf{Removal Rate}} & \textbf{Speed} & \multicolumn{1}{c|}{\textbf{Removal Rate}} & \textbf{Speed} & \multicolumn{1}{c|}{\textbf{Removal Rate}} & \textbf{Speed} \\ 
Weebly & 7031 & \multicolumn{1}{c|}{58.56\%} & 01:39 & \multicolumn{1}{c|}{20.65\%} & 04:41 & \multicolumn{1}{c|}{11.74\%} & 07:16 & \multicolumn{1}{c|}{13.12\%} & 5:38 & \multicolumn{1}{c|}{60.13\%} & 0:30 & \multicolumn{1}{c|}{23.46\%} & 7.13 \\ 
000webhost & 5934 & \multicolumn{1}{c|}{59.04\%} & 00:45 & \multicolumn{1}{c|}{13.82\%} & 07:23 & \multicolumn{1}{c|}{13.88\%} & 05:16 & \multicolumn{1}{c|}{10.70\%} & 4:10 & \multicolumn{1}{c|}{67.98 \%} & 4:02 & \multicolumn{1}{c|}{33.78\%} & 4.75 \\ 
Blogspot & 3156 & \multicolumn{1}{c|}{8.52\%} & 06:51 & \multicolumn{1}{c|}{25.12\%} & 07:03 & \multicolumn{1}{c|}{9.12\%} & 05:00 & \multicolumn{1}{c|}{11.10\%} & 3:57 & \multicolumn{1}{c|}{22.34\%} & 9:12 & \multicolumn{1}{c|}{30.11\%} & 4.07 \\ 
Wix.com & 2338 & \multicolumn{1}{c|}{64.55\%} & 02:16 & \multicolumn{1}{c|}{35.77\%} & 04:35 & \multicolumn{1}{c|}{12.73\%} & 01:29 & \multicolumn{1}{c|}{35.94\%} & 1:26 & \multicolumn{1}{c|}{43.66\%} & 4:18 & \multicolumn{1}{c|}{30.63\%} & 5.08 \\ 
Google Sites & 2247 & \multicolumn{1}{c|}{7.76\%} & 12:22 & \multicolumn{1}{c|}{28.45\%} & 18:08 & \multicolumn{1}{c|}{3.23\%} & 15:43 & \multicolumn{1}{c|}{5.28\%} & 22:14 & \multicolumn{1}{c|}{24.98\%} & 13:55 & \multicolumn{1}{c|}{14.40\%} & 16.8 \\ 
github.io & 942 & \multicolumn{1}{c|}{9.16\%} & 20:34 & \multicolumn{1}{c|}{21.46\%} & 07:05 & \multicolumn{1}{c|}{0.57\%} & 06:01 & \multicolumn{1}{c|}{13.06\%} & 15:52 & \multicolumn{1}{c|}{58.14\%} & 7:40 & \multicolumn{1}{c|}{20.44\%} & 12.5 \\ 
Firebase & 1416 & \multicolumn{1}{c|}{7.22\%} & 14:15 & \multicolumn{1}{c|}{26.86\%} & 09:09 & \multicolumn{1}{c|}{9.40\%} & 14:35 & \multicolumn{1}{c|}{12.09\%} & 10:41 & \multicolumn{1}{c|}{42.72\%} & 3:13 & \multicolumn{1}{c|}{26.08\%} & 11.5 \\ 
Squareup & 1736 & \multicolumn{1}{c|}{18.75\%} & 10:11 & \multicolumn{1}{c|}{34.45\%} & 10:58 & \multicolumn{1}{c|}{8.64\%} & 13:50 & \multicolumn{1}{c|}{6.68\%} & 14:48 & \multicolumn{1}{c|}{46\%} & 11:01 & \multicolumn{1}{c|}{34.22\%} & 19.32 \\ 
Zoho Forms & 498 & \multicolumn{1}{c|}{24.57\%} & 07:11 & \multicolumn{1}{c|}{15.77\%} & 10:30 & \multicolumn{1}{c|}{1.62\%} & 10:24 & \multicolumn{1}{c|}{8.84\%} & 10:12 & \multicolumn{1}{c|}{63.8\%} & 3:59 & \multicolumn{1}{c|}{31.22\%} & 14.56 \\ 
Wordpress & 786 & \multicolumn{1}{c|}{5.09\%} & 20:50 & \multicolumn{1}{c|}{28.96\%} & 17:07 & \multicolumn{1}{c|}{14.14\%} & 13:48 & \multicolumn{1}{c|}{8.18\%} & 47:28 & \multicolumn{1}{c|}{10.98\%} & 14:22 & \multicolumn{1}{c|}{12.47\%} & 19.95 \\ 
Google Forms & 1397 & \multicolumn{1}{c|}{11.96\%} & 06:17 & \multicolumn{1}{c|}{22.56\%} & 31:27 & \multicolumn{1}{c|}{3.87\%} & 07:37 & \multicolumn{1}{c|}{7.59\%} & 29:19 & \multicolumn{1}{c|}{39.45\%} & 4:26 & \multicolumn{1}{c|}{22.52\%} & 11.8 \\ 
Sharepoint & 2181 & \multicolumn{1}{c|}{7.64\%} & 05:07 & \multicolumn{1}{c|}{19.16\%} & 07:41 & \multicolumn{1}{c|}{13.73\%} & 01:37 & \multicolumn{1}{c|}{8.30\%} & 16:28 & \multicolumn{1}{c|}{16.65\%} & 2:08 & \multicolumn{1}{c|}{20.37\%} & 5 \\ 
Yolasite & 601 & \multicolumn{1}{c|}{7.52\%} & 07:05 & \multicolumn{1}{c|}{4.79\%} & 20:37 & \multicolumn{1}{c|}{10.46\%} & 13:28 & \multicolumn{1}{c|}{0} & N/A & \multicolumn{1}{c|}{24.22\%} & 1:31 & \multicolumn{1}{c|}{0} & N/A \\ 
GoDaddySites & 418 & \multicolumn{1}{c|}{5.84\%} & 04:58 & \multicolumn{1}{c|}{16.81\%} & 33:55 & \multicolumn{1}{c|}{0} & N/A & \multicolumn{1}{c|}{2.45\%} & 12:12 & \multicolumn{1}{c|}{32.85\%} & 11:44 & \multicolumn{1}{c|}{0} & N/A \\ 
MailChimp & 183 & \multicolumn{1}{c|}{23.67\%} & 18:11 & \multicolumn{1}{c|}{22.89\%} & 48:07 & \multicolumn{1}{c|}{2.15\%} & 08:16 & \multicolumn{1}{c|}{6.52\%} & 7:02 & \multicolumn{1}{c|}{21.34\%} & 5:19 & \multicolumn{1}{c|}{12.41\%} & 7.26 \\ 
glitch.me & 480 & \multicolumn{1}{c|}{21.31\%} & 34:47 & \multicolumn{1}{c|}{0} & N/A & \multicolumn{1}{c|}{3.10\%} & 10:33 & \multicolumn{1}{c|}{7.08\%} & 9:14 & \multicolumn{1}{c|}{11.67\%} & 16:48 & \multicolumn{1}{c|}{0.00\%} & 0 \\ 
hpage & 61 & \multicolumn{1}{c|}{19.60\%} & 11:45 & \multicolumn{1}{c|}{0} & N/A & \multicolumn{1}{c|}{0} & N/A & \multicolumn{1}{c|}{0} & N/A & \multicolumn{1}{c|}{13.11\%} & 21:27 & \multicolumn{1}{c|}{0} & 0 \\ \hline
\end{tabular}}
\end{table*}
\textbf{Low blocklist coverage and response time:}  
Table~\ref{table-fhd-overall} illustrates the blocklisting capabilities of PhishTank, OpenPhish, Google Safe Browsing (GSB), and eCrimeX against both FWB and regular (self-hosted) phishing URLs.
GSB covered 18.4\% of all FWB phishing URLs while having a median response time of 6 hours, which is much lower than their performance against self-hosted phishing attacks (74.2\% coverage and median response time of 51 minutes). On the other hand, the two open-source blocklists, PhishTank and Openphish performed the worst against FWB-based attacks by only being able to cover 4.1\% and 11.7\% of all URLs, respectively, while having a median response time of 7 hours and 13 hours, respectively. As was the case with GSB, the performance of both these open source blocklists was significantly better against self-hosted phishing URLs, a trend also seen from eCrimeX's performance, covering only 33\% of all FWB URLs, and having a median response time of nearly 9 hours. 
To look closer at their blocklisting capabilities, we measured their coverage rate over regular intervals up to a week after their first appearance in our dataset, illustrated in Figure~\ref{fig:blocklist_blocking_overtime}. 
We find that GSB covered nearly 60\% of all self-hosted phishing URLs within the first 3 hours of their appearance compared to only 11\% of FWB-hosted URLs during the same period. 
Within 24 hours, GSB could block up to 83\% of all self hosted phishing URLs while only detecting about 31\% of FWB phishing URLs. 
On the other hand, eCrimeX had similar coverage for self hosted phishing attacks (11\%) and FWB phishing URLs (8\%) at the 3 hours mark. The gap is significantly widened near the 24-hour mark, with eCrimeX covering 38\% of self-hosted phishing URLs, compared to only 13\% of FWB URLs. 
We also found that all four blocklists had less than 10\% coverage for nearly 40\% of all FWB URLs. In fact, only GSB has nearly 40\% coverage of 80\% of the FWB URLs, with the same statistic being only around 8\% for PhishTank, 12\% for OpenPhish, and 30\% for eCrimeX. Thus our measurement indicates that attackers have a lower chance of having their phishing attacks detected when created (and hosted) using an FWB, and even if the website does get detected, it would most likely take a longer time than self-hosted phishing attacks. 

\textbf{Blocklist coverage and response time varies on a per domain basis:}
\label{section-blocklist-per-domain} 
As evident from our findings in the previous section, detection efforts towards FWB phishing attacks were poor across all blocklists that we tested. We then investigated whether blocklisting varied on a per FWB service basis.
Table~\ref{table-fhd-per-domain} illustrates blocklists' coverage and response time over each of the FWBs found in our dataset. 
Our results indicate disproportionate performance by the blocklists based on which FWB the phishing website was created/hosted on. 
Interestingly, Weebly, 000webhost, and Wix, who collectively contributed to more than 48\%  of all URLs in our database, had higher coverage from all the blocklists when compared to phishing websites created on other FWBs. 
For example, GSB was able to detect 60.13\% of all URLs hosted on Weebly and had a median response time of 30 mins, compared to a less popular FWB Yolasite, whose URLs have a coverage of only 24.22\% with a median response time of 91 mins. 
However, an exception to this rule is phishing websites hosted on Blogspot, Google Sites, Google Forms, and Sharepoint, which despite being popular FWBs for hosting phishing attacks (collectively contributing to 28.5\% of our dataset), had low coverage and high response time over different blocklists. 
For example, while PhishTank shows a relatively poor performance across all popular FWBs, it is particularly evident in the case of phishing attacks hosted on Google Sites, where they covered only 3.23\% of the URLs and had a median response time of 15 hrs and 43 mins. We explore the reasons these attacks hosted on these FWBs have both low coverage and high response time at the end of our measurement study in Section~\ref{evasive-fhd}. Blocklists also had negligible or no coverage for some less popular FWBs. 
For example, PhishTank had no coverage for phishing websites created on \texttt{GoDaddySites} and \texttt{hpage}, while having negligible coverage for \texttt{github.io} (0.57\%), \texttt{Zoho Forms} (1.62\%), \texttt{Mailchimp} (2.1\%) and \texttt{glitch.me} (3.1\%). 
A similar trend was also observed in OpenPhish and eCrimeX. While GSB covers URLs from all FWBs, it also has low coverage for several of them.  
The lack of blocklist coverage for a particular FWB might entice attackers to more frequently abuse that service.


\begin{figure}[t]

\centerline{\includegraphics[width=0.75\columnwidth]{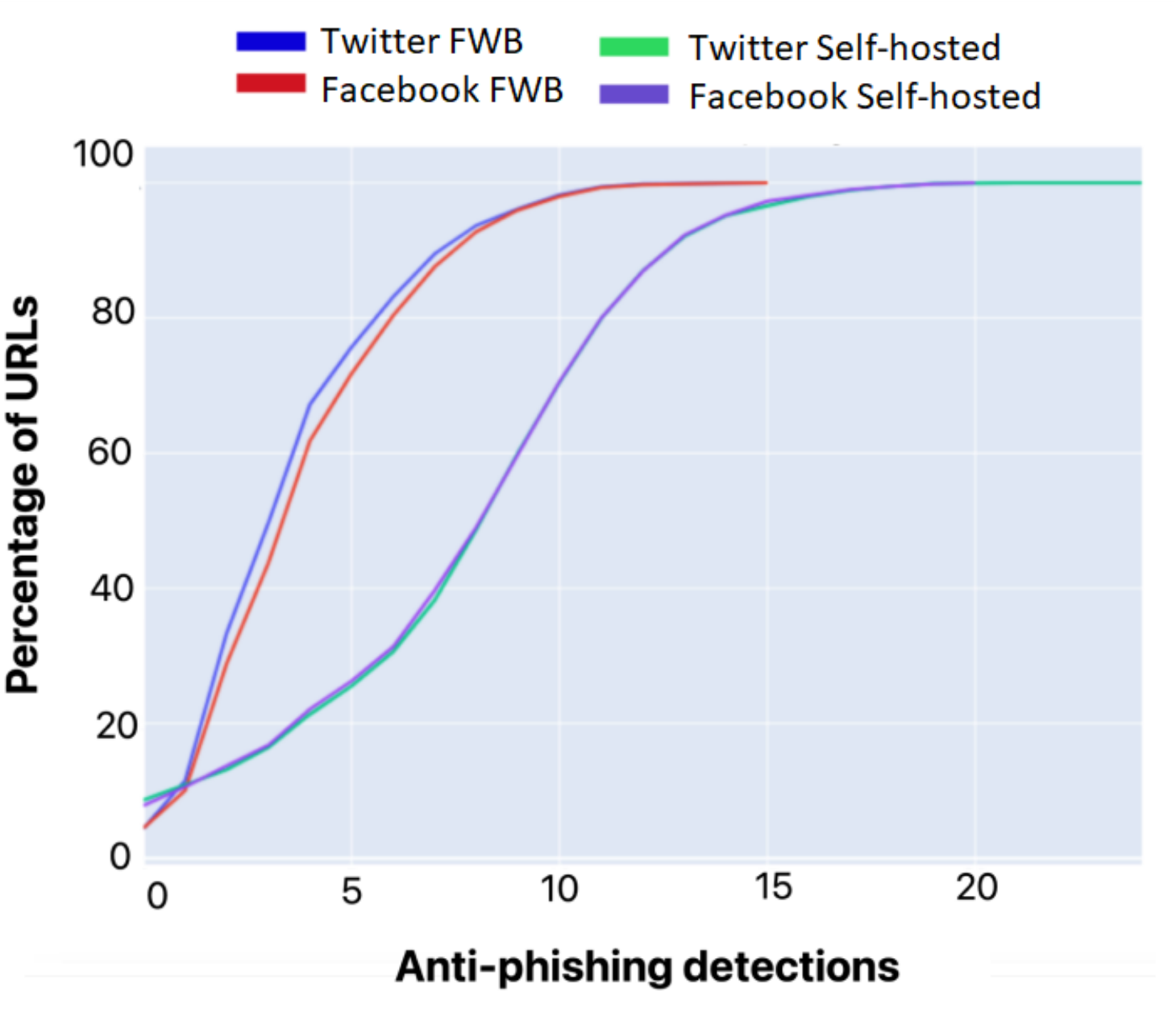}}
\caption{Cumulative distribution of antiphishing detections for FWB and self-hosted phishing websites}
\label{fig:antiphishing-cumul}
\end{figure}

\begin{figure*}[t!]
\centerline{\includegraphics[width=0.98\textwidth]{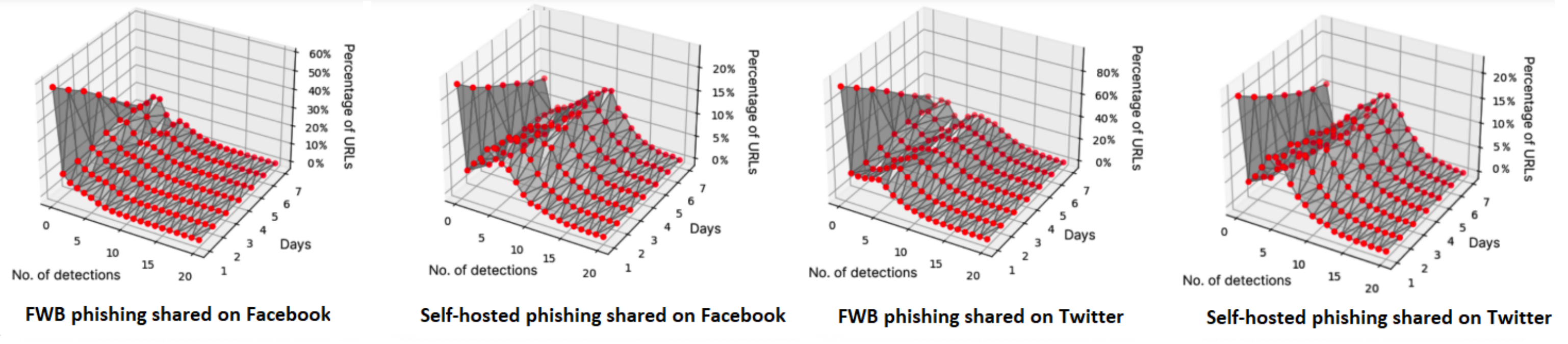}}
\caption{The percentage of FWB-based phishing URLs detected by anti-phishing engines over seven days.}
\label{fig:antiphishing-3d} 
\end{figure*}



\vspace{-10pt}
\subsection{Browser Protection Tool Coverage}
\label{antiphishingtool-coverage}
We conducted a longitudinal study to evaluate the effectiveness of 76 third-party anti-phishing tools on our dataset. These tools are included in security software or are available for download as browser extensions~\cite{mcafeewebadvisor,norton,bitdefender}.
Figure~\ref{fig:antiphishing-cumul} presents the distribution of anti-phishing detection for FWB-based URLs in comparison to self-hosted phishing URLs one week after their initial appearance on the respective social networks.
Our findings reveal that approximately 50\% of FWB attacks disseminated via these two social media platforms had a median detection of four tool detections, while self-hosted phishing attacks boasted a median of nine detections, indicating that FWB attacks are less frequently detected than their self-hosted counterparts. We find that FWB URLs, regardless of the social media platform share a similar tool detection trajectory, a trend that is also observed for self-hosted phishing URLs. 
We also compared the proportion of phishing attacks detected with the number of detections on a daily basis up to a week following their initial appearance, illustrated in Figure~\ref{fig:antiphishing-3d}. We discovered that nearly 75\% of FWB-based URLs posted on Twitter had only 2 detections (the minimum threshold for inclusion in our dataset), and after one week, 41\% of these URLs still  had four detections or fewer.
By contrast, 82\% of FWB URLs shared on Facebook had two detections on the first day,  43\% only had four detections or fewer after a week. 32\% of self-hosted phishing URLs shared via Twitter had 2 detections after one day, and 11\% of these URLs had four or fewer detections after a week. For Facebook URLs, 34\% of self-hosted phishing URLs had 2 detections after one day, and 8\% of these URLs had four detections or fewer after a week. While the graph provides more detailed insights, one key takeaway is that over time, FWB phishing URLs accrue fewer detections from anti-phishing tools than self-hosted phishing URLs.

\begin{figure}[t]
\centering
  \includegraphics[width=1\columnwidth]{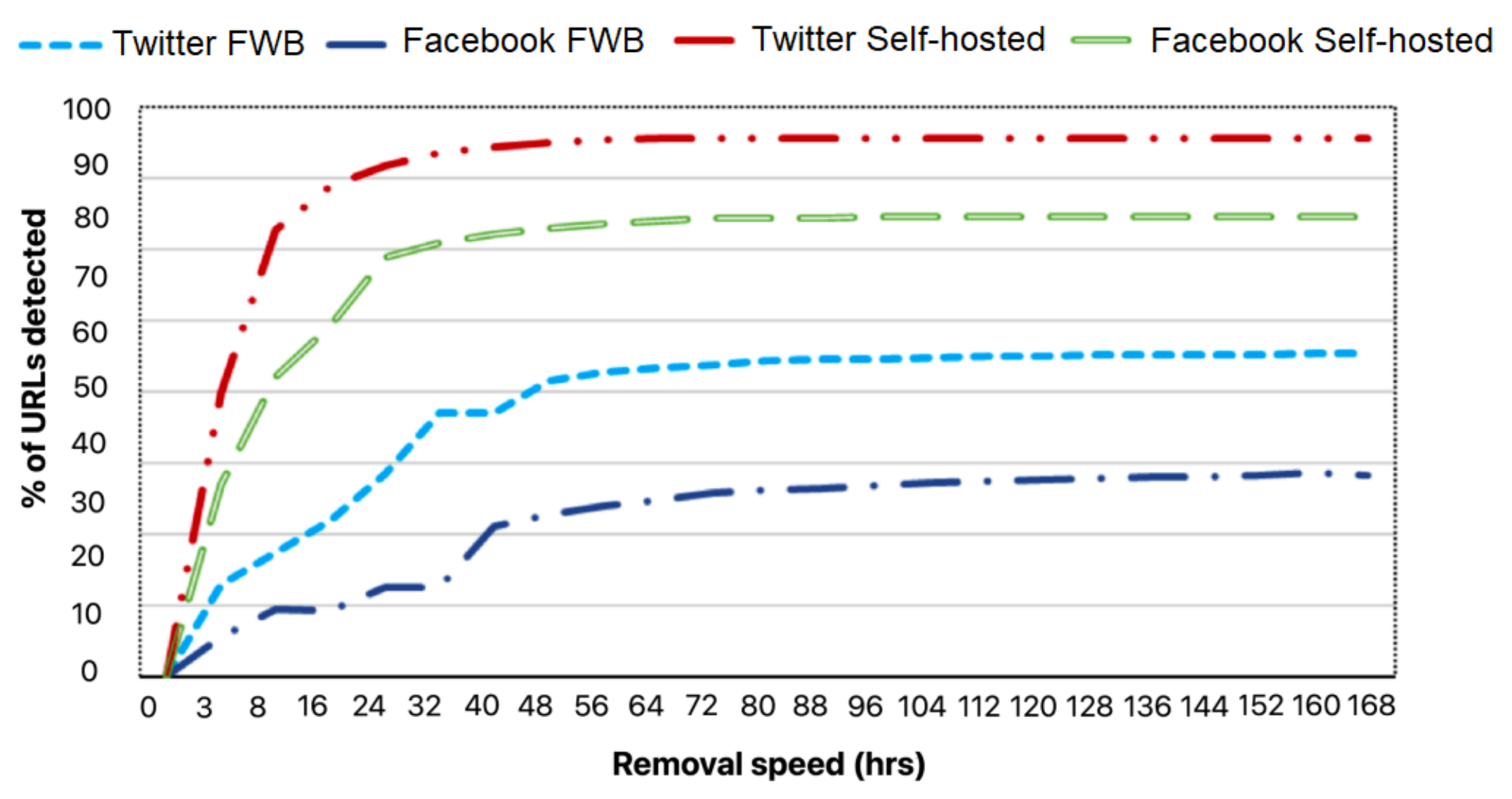}

\caption{Coverage and speed of platforms against FWB and self-hosted phishing attacks}
    \label{fig:platform-blocking-overtime}

\end{figure}

\subsection{Coverage and Response of FWB Services}
\label{domain-coverage}
Considering the low blocklist coverage of FWB phishing attacks, proactive removal of these websites is vital to prevent potential victims from visiting these websites. This action is usually taken by the domain registrar. 
As illustrated in Table~\ref{table-fhd-overall}, we found only 29\% of the websites were removed by the respective FWB service after two weeks since they first appeared in our dataset, at a median speed of 9 hours and 43 mins.
Table~\ref{table-fhd-per-domain} presents the coverage and response time of phishing attacks on a per FWB basis.
Weebly and 000webhost having the highest number of reported URLs also exhibit relatively high domain removal rates, 58.56\% and 59.04\%, respectively, suggesting that these FWBs are heavily targeted by phishing attacks, prompting more proactive measures. 
Their domain response speeds are also relatively quick with a median speed of 1 hr 39 mins and 45 mins respectively, indicating efficient phishing attack mitigation processes.
Wix.com, despite having fewer URLs than Weebly and 000webhost, boasts the highest domain removal rate at 64.55\%, showcasing its effectiveness in dealing with such threats. Their response speed, however, is longer than Weebly and 000webhost but still faster than most other FWBs.

FWBs such as Blogspot, Google Sites, Github.io, Firebase, Sharepoint, Yolasite, and GoDaddySites have lower domain removal rates. This could point to less effective handling of phishing attacks. Interestingly, their response speeds vary widely, with Github.io and Firebase having particularly long response times with median removal times of 14 hrs 15 mins and 20 hrs 34 mins respectively , perhaps indicating less efficient processes for dealing with such threats. 
On the other hand, Squareup, Zoho Forms, Google Forms, MailChimp, glitch.me, and hpage exhibit a moderate domain removal rate. Their response speeds are also quite varied, suggesting differences in their effectiveness in dealing with phishing attacks.
Thus, our findings indicate that the handling of phishing attacks varies widely among FWBs, with some services demonstrating a strong commitment to mitigating threats that are created on their platform while others lag behind. This variation can potentially have a significant impact on the overall security of these domains, and potentially lead to attackers favoring FWBs which have a lower response. 
\textbf{Response to reporting:}  
There are several factors that can influence the coverage and removal speed of phishing attacks by FWBs. Some of these are discussed towards the end of this Section in~\ref{evasive-fhd}. However, one factor that might significantly impact their effectiveness is how each FWD and registrar responds to our reports.
WordPress, GoDaddySites, Firebase, Google Sites, Sharepoint, and Yolasite did not respond to any of our reports, which raises concerns about their commitment to dealing with such issues. On the other hand, Squareup (23.7\%), Github.io (37.4\%), Google Sites (15.2\%), and Blogspot (28.3\%) acknowledged some of the initial reports, creating a support ticket but did not follow up with a resolution. Values inside parentheses denote the percentage of URLs where we received an initial response. 
Incidentally, these FWBs had the lowest coverage rates and longest response times, as evident from our findings in Table~\ref{table-fhd-per-domain}. This trend raises questions about whether they have implemented suitable measures to validate and respond to phishing URLs reported by our framework. On the contrary, FWBs with high coverage rates, such as Weebly (71.6\%), Wix (65.3\%), 000webhost (82.7\%), and Zoho Forms (70.4\%), were very responsive to our reports and followed up with additional information and the subsequent removal of both the website and user account associated with the phishing attack. Values inside parentheses denote the percentage of URLs where we received an initial response and a follow-up. 
Given that these FWBs are frequently targeted by phishing attacks, they seem to adopt a more alert stance towards these threats, emphasizing rapid action for their removal.
Thus, the responsiveness of FWBs to these reports can significantly impact their coverage and response speed, affecting the overall security of these platforms.
\vspace{-8pt}
\subsection{Platform Effectiveness}
\label{platform-coverage}
By tracking the activity of posts that shared the URLs over the course of a week, we found that both of the platforms, i.e., Twitter and Facebook, collectively cover about 23\% of all FWB URLs, in comparison to 71.9\% of self-hosted phishing URLs.
The possibility exists that the tweets may have been deleted by the users themselves, not the platforms. However, prior research~\cite{li2014towards,bauer2013post} has shown that less than 2\% of tweets are typically removed on social media platforms, translating to minimal noise in our measurement data.

The second column of Table~\ref{table-fhd-per-domain} displays the platforms' coverage rate and response time to FWB attacks per FWB service basis. Unlike anti-phishing blocklists, which vary in their proficiency in detecting certain domains, or domain providers who demonstrate varying efficiency in removing FWB URLs, platform coverage (though low) does not differ significantly between FWBs. 
We do observe a higher median response time for popular FWBs, like Google Sites and Forms (18.08 hours and 31.27 hours, respectively), WordPress (17.07 hours), and Sharepoint (7.41 hours). Conversely, phishing websites hosted on less popular FWBs, such as Yolasite (20.37 hours), glitch.me (not applicable), GoDaddySites (33.55 hours), and MailChimp (48.07 hours) take longer to remove.

To illustrate the coverage and removal speed of FWB URLs shared on Twitter at a granular level, we monitored the coverage rate of these URLs over multi-hour periods up to a week (168 hours) after their initial appearance in our dataset. This is displayed in Figure~\ref{fig:platform-blocking-overtime}. 
Within the first three hours, Twitter and Facebook removed only about 10\% and  6\% of FWB-based phishing URLs shared on their respective platforms, compared to the removal of 32\% and 47\% of self-hosted phishing URLs within the same time frame. After 16 hours, Twitter managed to remove over 70\% of self-hosted phishing URLs while removing only 21\% of FWB attacks. In the same period, Facebook removed over half of the self-hosted phishing URLs but only 10\% of FWB URLs.
Thus, phishing attacks hosted using FWBs on Twitter and Facebook tend to persist for much longer than self-hosted attacks. 

\begin{figure}[t]
\begin{center}
\includegraphics[width=0.8\columnwidth]{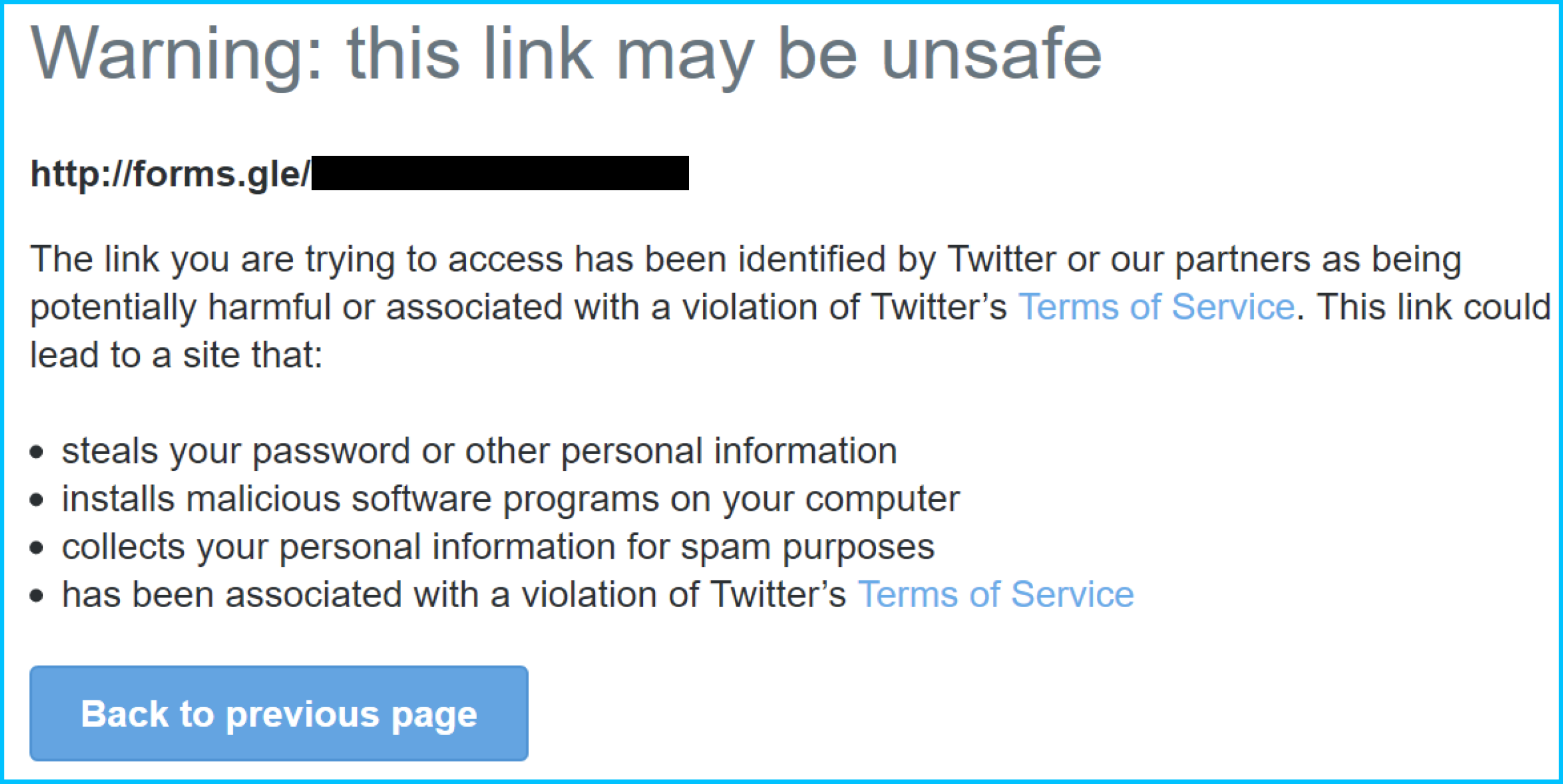}
 \caption{User alert displayed when attempting to access a website flagged as malicious by Twitter}
\label{twitter_warning_msg}
\end{center}
\end{figure}

From the end-users point of view, Figure~\ref{twitter_warning_msg} showcases the warning page Twitter used to display when a user attempted to navigate to a known malicious site. However, since Twitter's transition to the brand "X" in July 2023~\cite{twitter_rebrand}, this warning mechanism seems to have been discontinued, even though the platform still actively removes posts featuring phishing URLs. Unlike Twitter, to the best of our knowledge, Facebook does not employ user-facing warning pages and instead directly deletes posts containing malicious links.

\subsection{Qualitative Analysis of Evasive Attacks}
\label{evasive-fhd}

Nearly 14.2\% of all URLs in our dataset (n=4,459 URLs) did  not contain credential fields.
Several of these URLs belonged to popular FWBs such as Google Sites, Google Forms, Blogpsot, and Sharepoint.
To investigate this issue closely, we randomly sampled 1K URLs from this set and qualitatively analyzed their content (both appearance and codebase) to identify the attack vectors, which led us to identify three variants of malicious websites that deviate from regular phishing strategies. We developed heuristics to automatically identify these attack vectors across our dataset's FWB phishing attacks.

\begin{figure}[t]
\centering
  \includegraphics[width=0.9\columnwidth]{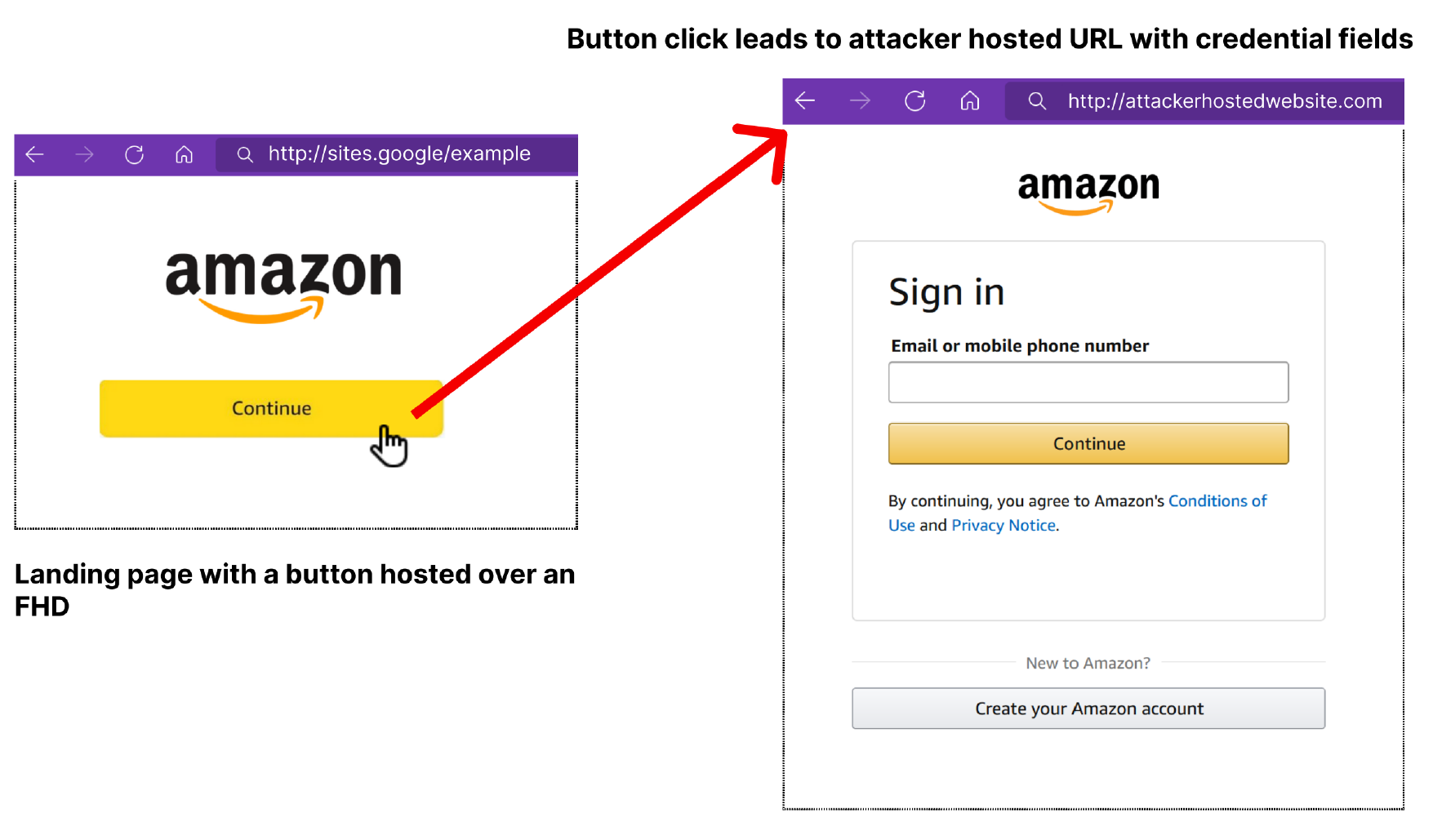}
\caption{Two-step phishing example: A landing page on hosted FWB features only a button. Clicking on the button redirects users to a separate, attacker-controlled site that asks for login details.}
  \label{fig:button_phish}

\end{figure}

\textbf{Linking to other phishing pages:} We found 539 (about 24\%) URLs hosted on Google Sites containing a button, clicking on which led to a phishing website on a different domain. 
This design serves two evasive purposes - Avoiding credential fields on the website,
to more effectively evade detection by anti-phishing bots, and - by not directly sharing the linked phishing page accessed through user clicks, attackers lower the risk of discovery and detection.
These URLs' low coverage and response time are evident from the fact that few approaches~\cite{liu2022inferring} have been able to identify this evasion tactic. 
For 174 URLs, the attackers used another FWB as its linked page, which can be more effective at evasion.
An example of this attack is illustrated in Figure~\ref{fig:button_phish}.
We also found 349 (about 16\%) URLs on Sharepoint, 293 (about 21\%) on Google Forms, and 447 (about 14\%) on Blogspot.

\begin{figure}[t]
\centering
  \includegraphics[width=0.7\columnwidth]{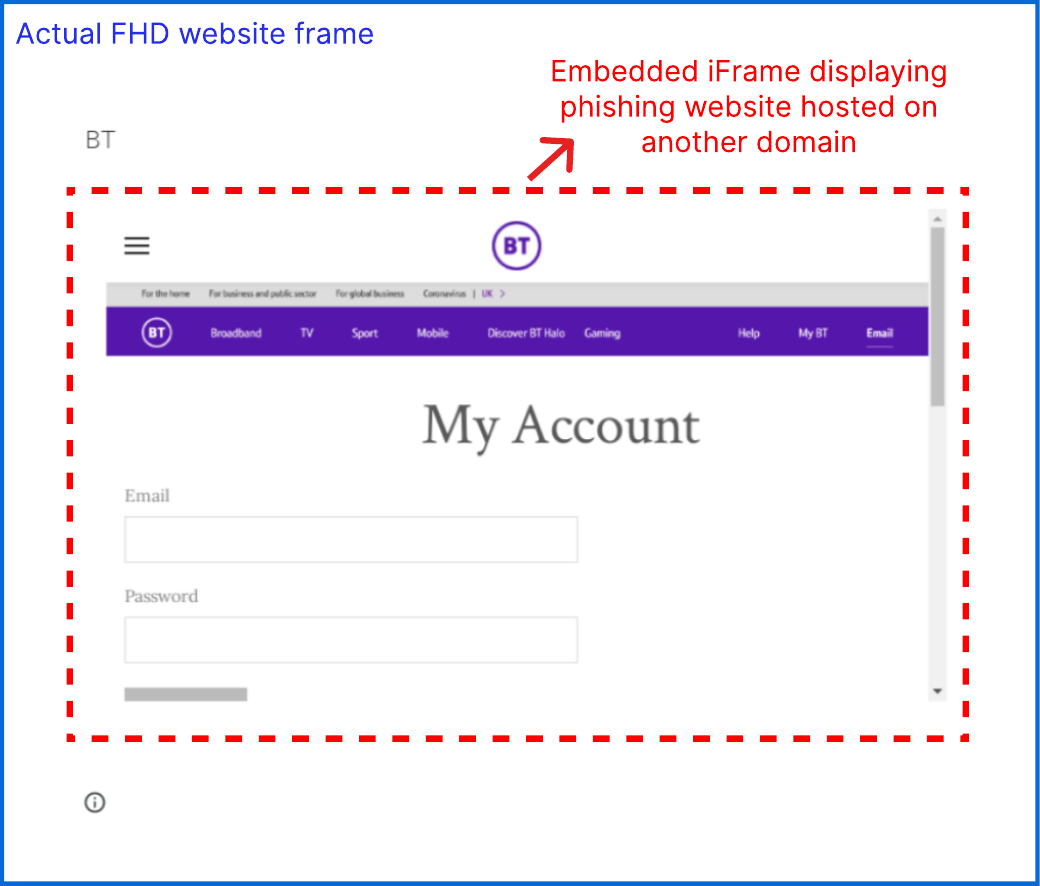}
\caption{FWB containing an embedded i-Frame for another attack on an external domain.}
  \label{fig:btb-fhd}

\end{figure}

\textbf{i-Frames Embedding External Attacks:} In 427 URLs on Google Sites and 473 URLs on Blogspot, attackers embedded a phishing site's i-frame into a benign website's code hosted on an FWB. This concealed i-Frame activates either by button click on the parent website or by loading alongside it. Figure~\ref{fig:btb-fhd} illustrates an example on Google Sites. These attacks could evade prevalent anti-phishing tools, as i-frame content is client-side rendered and unavailable to tools scanning for malicious code~\cite{oest2020phishtime}. We discovered these FWB domains comprise 62\% of all i-frame phishing attacks in our dataset.

\textbf{Malicious drive-by downloads:} 725 (23\%) of the URLs on Blogspot, 651 URLs (29\%) on Google Sites, and 1,177 URLs (54\%) on Sharepoint were used to share malicious downloads hosted on third-party websites. 
For each URL that triggered a download, we scanned the file using VirusTotal.
If the file had four or more detections (a threshold proposed in literature~\cite{oprea2018made, sharif2018predicting}), the FWB URL was marked as distributing malicious drive-by downloads. 
Combining this with the absence of credential-requiring fields in such websites hosting malicious downloads might further indicate the low coverage of these attacks from anti-phishing entities. 
Interestingly, 741 malicious URLs were hosted on Sharepoint, mimicking OneDrive or Office 365 documents, highlighting targetted spoofing of Microsoft Office products.

\section{Related Work} 
\textbf{Anti-phishing systems:} Phishing attacks pose a significant threat to users~\cite{junger2017priming,bullee2018anatomy}, prompting the integration of anti-phishing tools~\cite{safebrowsing,mcafeewebadvisor,bitdefender} in web-browsers. These tools rely on blocklists, databases updated with known phishing sites taken from community reports~\cite{phishtank}, third-party data~\cite{openphish}, or web crawler frameworks~\cite{oest2020phishtime}. Researchers have improved phishing detection using automated models incorporating machine and deep learning~\cite{sahingoz2019machine,rao2019detection}, and heuristics~\cite{nguyen2013detecting,sreedharan2016systems}.

Nevertheless, recent studies have highlighted the vulnerability of blocklists and automated tools to evasive attacks. For instance, Oest et al.\cite{oest2020phishtime} unveiled the significant impact of server-side cloaking attacks on blocklist detection, while Acharya et al.\cite{acharya2021phishprint} and AlEroud et al.~\cite{aleroud2020bypassing} outlined ways attackers use browser-based fingerprinting obfuscations and meta-URL features to bypass automated anti-phishing crawlers.

Our work explores how attackers leverage features of several Free Website Builder services to evade detection. We characterize unique aspects of these attacks, enabling the creation of a real-time automated machine learning model for detecting such threats in the wild.

\textbf{Anti-phishing measurement}: Anti-phishing measurements at scale are instrumental in revealing emerging threats and gaps in the current detection ecosystem. Oest et al's frameworks, PhishTime~\cite{oest2020phishtime} and Golden Hour~\cite{oest2020sunrise}, focus on real-time phishing threat identification and traffic monitoring, stressing the need for fortified mobile security and evidence-based reporting protocols. On another front, Bijman et al~\cite{bijmans2021catching} sheds light on the influence of off-the-shelf phishing kits in the Dutch financial sector and suggested policy enhancements. Further, Bell's et al~\cite{bell2020analysis} examination of three phishing blacklists underscores the necessity for improved blacklist management, given their varying size, update frequency, and potential effectiveness. Our framework, FreePhish contributes to this collective effort by identifying and measuring the impact and efficacy of phishing websites created using Free Website Building services against anti-phishing tools, blocklists, hosting services, and sharing platforms.

\textbf{Phishing Attack Costs:} Phishing kits \cite{proofpointphishingkit} have reduced the technical expertise needed to execute phishing attacks as they can automate the creation and hosting of phishing sites and can incorporate evasion strategies\cite{oest2018inside,bijmans2021catching}. However, they are typically costly for attackers~\cite{herley2010nobody}, and free versions often contain malware~\cite{cova2008there}. Furthermore, attackers typically acquire inexpensive web domains with non-traditional top-level domains (TLDs) from hosting services, such as \textit{.store}, \textit{.top}, or \textit{.live}\cite{gupta2014emerging}, given the short lifespan of most phishing sites\cite{oest2020sunrise}. Yet, the high volume of phishing attacks on these TLDs invites more scrutiny from anti-phishing blocklists and hosting services~\cite{detectionagainsttld:2022,agten2015seven,verma2015character}, thereby reducing the attacks' uptime. On the other hand, our research highlights how attackers leverage Free Website Builders to alleviate the financial burdens typically associated with phishing attacks. These platforms offer a host of advantages, including premium top-level domains, instant SSL certification, and user-friendly design interfaces, thereby making it possible to host sophisticated phishing sites at no cost.
\section{Conclusion}
\label{conclusion}
\begin{figure}[t]
\centering
  \includegraphics[width=0.9\columnwidth]{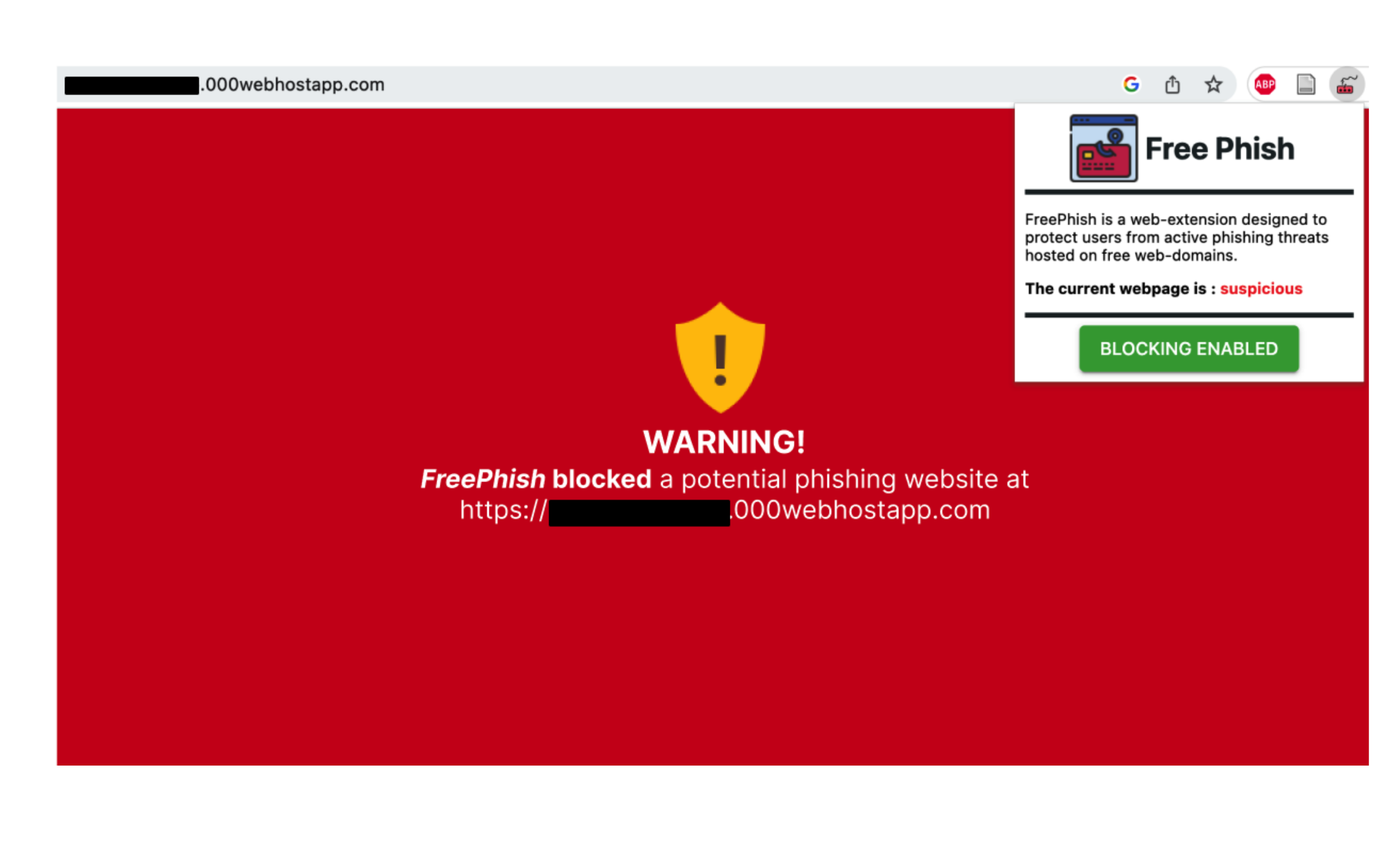}

\caption{The FreePhish web-extension in action}
  \label{fig:extension-in-action}
\end{figure}

Our study investigated and described the tactics used to construct phishing websites using free website builders. We discovered that attackers exploit various features offered by these services to create phishing websites that can evade detection by anti-phishing entities. To combat this, we developed a framework, \textit{FreePhish}, designed to monitor and promptly identify FWB phishing attacks shared on social media websites. This framework reports these attacks for immediate takedown and assesses the efficiency of anti-phishing entities in countering these threats. In six months, we identified and reported over 31.4K newly generated FWB phishing URLs, thus providing both FWB services and social media platforms a tool to mitigate and remove these threats more effectively. We also FreePhish as a free web extension that can be installed on Chromium-based browsers to protect users from accessing FWB phishing URLs. 
Figure~\ref{fig:extension-in-action} illustrates the extension in action where it has blocked a malicious FWB phishing attack.
\urlstyle{tt}
The web extension and source code of the FreePhish framework source code is available at: \url{https://github.com/UTA-SPRLab/freephish/}.

\section{Ethics}

To help the research community to evaluate the phishing attacks examined in our study, but also considering the sensitivity and potential risk of these threats , we open-source a sample of our dataset ($\sim$28\% of all URLs used in our initial dataset) that have been confirmed to have been removed by their respective hosting provider at \url{https://github.com/UTA-SPRLab/freephish/}
\urlstyle{rm}. After finishing our measurement study, we have also reported our URLs to the APWG eCrimeX blocklist~\cite{ecrimex:2022}, and plan to continue reporting  FWB phishing attacks found by our FreePhish framework to the respective FWBs and social media platforms.
The qualitative analysis of this data was conducted securely, using a designated computing system situated within our research laboratory. The two coders involved had access to the URLs strictly within this controlled environment. We have also not retained or made use of any personally identifiable information from the posts gathered from Twitter and Facebook. The data has been used exclusively in an aggregated manner, specifically for relevant analysis within the scope of our work.

{\footnotesize \bibliographystyle{acm}
\bibliography{refs}}
\appendix

\section{Code similarity between FWB websites}
\label{appendix-code-similarity}

In Section~\ref{characterization}, we calculate the code similarity between phishing attacks and legitimate websites that are created on FWBs. 
To calculate the similarity between two websites (Website A and Website B), we first extracted the tag elements from each website. For each tag element \(T\) in Website \(A\), we calculated its Levenshtein distance (LV) to every tag element in Website \(B\). The tag from Website \(B\) with the lowest Levenshtein distance was considered the most similar to the \(T\) tag of Website \(A\). This is denoted as \(T_{\text{max}}\). Thus,
\[ T_{\text{max}} = \min \{ \text{LV}(T, T_B) \ | \ \text{for each}\ T_B \in \text{Website B} \} \]

We then calculated the similarity between Website \(A\) and Website \(B\) (\(\text{sim}_{A \text{ to } B}\)) as the median of all \(T_{\text{max}}\) values.

\[ \text{sim}_{A \text{ to } B} = \text{median}(T_{\text{max}} \ | \ \text{for all}\ T \in \text{Website A}) \]

We also calculated the similarity from Website \(B\) to Website \(A\) (\(\text{sim}_{B \text{ to } A}\)) using the same procedure.

\[ \text{sim}_{B \text{ to } A} = \text{median}(T_{\text{max}} \ | \ \text{for all}\ T \in \text{Website B}) \]

Finally, the similarity between the two websites (\(\text{sim}_{A,B}\)) was calculated as the mean of \(\text{sim}_{A \text{ to } B}\) and \(\text{sim}_{B \text{ to } A}\).

\[ \text{sim}_{A,B} = \text{mean}(\text{sim}_{A \text{ to } B}, \text{sim}_{B \text{ to } A}) \]

\end{document}